\documentclass[12pt]{article}
\usepackage{hyperref}
\usepackage{amsmath}
\numberwithin{equation}{section}

\begin{document}

\newcommand{\be}{\begin{equation}}
\newcommand{\ee}{\end{equation}}
\newcommand{\ber}{\begin{eqnarray}}
\newcommand{\eer}{\end{eqnarray}}


%
\title{Hamiltonian cosmology in bigravity and massive gravity}

\author{{Vladimir O. Soloviev~\footnote{E-mail:Vladimir.Soloviev@ihep.ru}}\\
{\small Institute for High Energy Physics,}\\
{\small National Research Center ``Kurchatov Institute''} }
%
%

%
\maketitle

\begin{abstract}
In the Hamiltonian language we provide a study 
of flat-space cosmology  
in bigravity and massive gravity constructed  mostly 
with de Rham, Gabadadze, Tolley (dRGT) potential. 
It is demonstrated that the Hamiltonian methods are powerful
not only in proving the absence of the Boulware-Deser ghost, but also in solving other problems. The purpose of this work is to give an introduction both to the Hamiltonian formalism  and to the cosmology of bigravity.
We 
sketch
three roads to the Hamiltonian of bigravity with the dRGT potential: the metric, the tetrad and the minisuperspace approaches. 
\end{abstract}

\newpage
\tableofcontents

\section{Introduction}
One of the actual problems in modern physics is the problem of dark energy. There are different approaches to it, and one of them is to revise the standard theory of gravitation, i.e. the General Relativity (GR). New predictions are expected 
for physics at the very large distances, and they may be provided by making gravity massive. The popular way to do so is to introduce a new 
tensor field with the same properties as the metric. From the aesthetic point of view it is natural to expect that this new field will be dynamical. 
Such a model is called bigravity or bimetric gravity.

The crucial problem in constructing the bigravity or massive gravity models is to avoid the Boulware-Deser ghost~\cite{Boulware:1973my}. This problem was solved by de Rham, Gabadadze, and Toley~\cite{deRham:2010ik,deRham:2010kj} (dRGT), who proposed the genuine form of potential involving the matrix square root. The proof that this theory is ghost-free has been given first  in the metric Hamiltonian approach~\cite{HR_bi,HR2}. But it is hard to work with the dRGT potential in metric formalism, whereas in tetrad approach the calculations become explicit. Fortunately, the background cosmology that will be discussed in this work exploits only rather simple form for both the two metric tensors, and in the so-called minisuperspace approach the relevant matrix is diagonal, so the square root of it can be found directly.  

Let us mention that the Boulware-Deser ghost appears below even in minisuperspace models for the two cases: first, when a non-dRGT potential is chosen (see Section 6), second, when one matter field minimally couples to a pair of metrics (see subsection 5.2).

A lot of work in cosmology of  bigravity has been already done by many authors~\footnote{We are sorry for citing only part of them and not trying to give a complete list of references.}. As a rule, the main instruments were the Lagrangian equations and Bianchi identities. 
The aim of this work is to demonstrate the main features and power of the Hamiltonian formalism in analyzing the variety of cosmological problems from a unique viewpoint. This variety arises due to many ways by which the matter fields can couple to the  gravitational fields. We consider the minimal interactions of two species of matter with the two metrics, the double interactions of one matter with the two metrics, and the effective metric interacting with the matter. We also illustrate the general scheme by an example of the non-dRGT potential. For simplicity we consider only flat space dynamics, and each metric has only one matter source. 

Here all the local physics is ignored including the important question of stability of the background cosmological solutions. This is discussed in many articles, for example, see recent publications~\cite{Comelli:2015pua,Gumrukcuoglu:2015}.  We are impressed by the statement~\cite{25March2015} that it is possible to cure the growing perturbations by taking a tiny Planck mass for the second metric $f_{\mu\nu}$.

In Sections 2 and 3 we briefly explicate the construction of the bigravity Hamiltonian (first, in metric variables; second, in tetrad variables), for a technically more detailed presentation we address the reader to articles~\cite{SolTch,SolTch2,Comelli,Sol,Soloviev:2014eea,Kluson_tetrad}. For approaches to the Hamiltonian formalism that are not discussed here see~\cite{HR_bi,HR2,Golovnev,Deser}. The title of this work should remind about a good old book~\cite{Ryan}. 
The mini-superspace Hamiltonians are constructed in Section 4, the machinery given there is applied to the concrete derivations of the cosmological equations in Sections 5, 6.

\section{
Metric variables}
The Lagrangian density of the bigravity theory may be written as follows
$$
{\cal L}={\cal L}_g+{\cal L}_f
+{\cal L}_M-\frac{2m^2}{\kappa}
U(f_{\mu\nu},g_{\mu\nu}),
$$
where the two copies of the GR Lagrangian are involved
$$
{\cal L}_{g}=\frac{1}{\kappa_g}\sqrt{- g}{g}^{\mu\nu}R^{(g)}_{\mu\nu},\qquad {\cal L}_{f}=\frac{1}{\kappa_f}\sqrt{- f}{f}^{\mu\nu}R^{(f)}_{\mu\nu}.
$$
The matter Lagrangian and the potential for  interaction of the  two metrics are also included in ${\cal L}$. For simplicity, as an example of matter, we will take the scalar field with a minimal interaction with some metric ${\cal G}_{\mu\nu}$, 
\be
{\cal L}_\phi=\sqrt{-{\cal G}}\left(-\frac{1}{2}{\cal G}^{\mu\nu}\phi_{,\mu}\phi_{,\nu}-{\cal U}(\phi) \right).
\ee
The role of  metric below may be played by $g_{\mu\nu}$, or by $f_{\mu\nu}$, or by their combination called the effective metric (see subsection 5.3). The constant $\kappa$ standing in front of the potential may be chosen coinciding with standard Newtonian one $\kappa_g=16\pi G$, but in some works it is different, for example,
\be
\frac{1}{\kappa}=\frac{1}{\kappa_g}+\frac{1}{\kappa_f},
\ee
so we start with the general notation.
We denote $3+1$ components of a general metric ${\cal G}_{\mu\nu}$ and  its inverse in the ADM coordinate basis as follows
\begin{eqnarray}
 ||{\cal G}_{\mu\nu}||&=&\left(\begin{array}{cc}-{\cal N}^2+\psi_{mn}{\cal N}^m{\cal N}^n & \psi_{jk}{\cal N}^k\\
\psi_{ik}{\cal N}^k & \psi_{ij}\end{array}\right),\label{eq:calG1}\\
||{\cal G}^{\mu\nu}||&=&\left(\begin{array}{cc}-{\cal N}^{-2} & {\cal N}^j{\cal N}^{-2}\\{\cal N}^i{\cal N}^{-2} &\psi^{ij}-{\cal N}^i{\cal N}^j{\cal N}^{-2}\end{array}\right),\label{eq:calG2}
\end{eqnarray}
where ${\cal N}$ is the lapse, ${\cal N}^i$ is the shift vector, $\psi_{ij}$ is the induced metric on  spatial hypersurfaces of fixed time, and $\psi^{ij}$ is the inverse matrix for it.
The interaction potential of the two metrics suggested by de Rham, Gabadadze and Tolley~\cite{deRham:2010ik,deRham:2010kj} is constructed as a linear combination of the symmetric polynomials $e_i$ of matrix $\mathsf{X}^\mu_\nu=\left(\sqrt{g^{-1}f}\right)^\mu_\nu$
\begin{eqnarray}
e_0&=&1,\nonumber\\
e_1&=&\lambda_1+\lambda_2+\lambda_3+\lambda_4,\nonumber\\
e_2&=&\lambda_1\lambda_2+\lambda_2\lambda_3+\lambda_3\lambda_4+\lambda_4\lambda_1+\lambda_1\lambda_3+\lambda_2\lambda_4,\nonumber\\
e_3&=&\lambda_1\lambda_2\lambda_3+\lambda_2\lambda_3\lambda_4+\lambda_1\lambda_3\lambda_4+\lambda_1\lambda_2\lambda_4,\nonumber\\
e_4&=&\lambda_1\lambda_2\lambda_3\lambda_4,\label{eq:e_i}
\end{eqnarray}
where $\lambda_i$ are eigenvalues of  $\mathsf{X}$. Then
$$
U=\sqrt{-g}\sum_{n=0}^4\beta_ne_n(\mathsf{X})=\beta_0\sqrt{-g}+\ldots+\beta_4\sqrt{-f}\equiv N\tilde U, 
$$

Our plan of constructing the Hamiltonian formalism is the following. First, we  import the canonical variables and the expressions of Hamiltonians from the two copies of General Relativity
\ber
H_g&=&\int d^3x\left( \bar N\bar{\cal H}+\bar N^i\bar {\cal H}_i\right),\\
H_f&=&\int d^3x\left( N{\cal H}+ N^i{\cal H}_i\right),
\eer
where
\ber
\bar{\cal H}&=& -\frac{1}{\sqrt{\gamma}}\left(\frac{1}{\kappa_g}\gamma R^{(3)}+\kappa_g\left(\frac{\pi^2}{2}-\mathrm{Tr}\pi^2 \right) \right)
,\label{eq:constraint1}         \\
\bar {\cal H}_i&=& -2\pi_{i|j}^j,\label{eq:constraint2} \\
{\cal H}&=& -\frac{1}{\sqrt{\eta}}\left(\frac{1}{\kappa_f}\eta R^{(\eta)}+\kappa_f\left(\frac{\Pi^2}{2}-\mathrm{Tr}\Pi^2 \right) \right)
\label{eq:constraint01}         ,\\
{\cal H}_i&=&  -2\Pi_{i|j}^j.
\label{eq:constraint02}    
\eer
The canonical gravitational variables have the standard Poisson brackets
\ber
\{\gamma_{ij}(x),\pi^{k\ell}(y)\}&=&\frac12 \left(\delta_i^k\delta_j^\ell+\delta_i^\ell\delta_j^k\right)\delta(x,y),\\
\{\eta_{ij}(x),\Pi^{k\ell}(y)\}&=&\frac12 \left(\delta_i^k\delta_j^\ell+\delta_i^\ell\delta_j^k\right)\delta(x,y).
\eer
The ADM-decompositions for the two space-time metrics are as follows
\begin{eqnarray}
 ||f_{\mu\nu}||&=&\left(\begin{array}{cc}-N^2+\eta_{mn}N^mN^n & \eta_{jk}N^k\\
\eta_{ik}N^k & \eta_{ij}\end{array}\right),\\ 
||f^{\mu\nu}||&=&\left(\begin{array}{cc}-N^{-2} & N^jN^{-2}\\N^iN^{-2} &\eta^{ij}-N^iN^jN^{-2}\end{array}\right),\\
|| g_{\mu\nu}||&=&\left(\begin{array}{cc}-{\bar N}^2+\gamma_{mn}{\bar N}^m{\bar N}^n & \gamma_{jk}N^k\\
\gamma_{ik}{\bar N}^k & \gamma_{ij}\end{array}\right),\\
 ||g^{\mu\nu}||&=&\left(\begin{array}{cc}-{\bar N}^{-2} & {\bar N}^j{\bar N}^{-2}\\
{\bar N}^i{\bar N}^{-2} &\gamma^{ij}-{\bar N}^i{\bar N}^j{\bar N}^{-2}\end{array}\right),
\end{eqnarray}
Next we  construct the matter Hamiltonian
$$
H_M=\int d^3x \left({\cal N}{\cal H}_M+{\cal N}^i{\cal H}_{Mi} \right),
$$
and in case the role of matter is played by one scalar field with a minimal coupling to one metric ${\cal G}_{\mu\nu}$, according to 
Eqs.(\ref{eq:calG1}), (\ref{eq:calG2}) we obtain
$$
{\cal H}_M=\frac{\pi_\phi^2}{2\sqrt{\psi}}+\frac{\sqrt{\psi}}{2}\psi^{ij}\partial_i\phi\partial_j\phi+\sqrt{\psi}{\cal U}(\phi).
$$
In dealing with two or even more metrics instead of the coordinate basis it is suitable to use the geometrical basis~\cite{Kuch1,Kuch2,Kuch3,Kuch4} formed by a unit normal to the state hypersurface $n^\alpha$ and three vectors tangential 
to this hypersurface $e^\alpha_i$. Let $X^\mu$ be an arbitrary space-time coordinate system, and $(t,x^i)$ is the ADM coordinate system, i.e. functions $X^\mu=X^\mu(t)$ give one-parametrical foliation of the space-time by spacelike hypersurfaces, whereas $x^i$ are internal coordinates on a hypersurface. Then $e^\mu_i=\partial X^\mu/\partial x^i$ are the three tangential vectors.
Of course, we need a space-time metric to construct the unit normal, let us suppose this metric to be $f_{\mu\nu}$, then
$$
f_{\mu\nu}n^\mu n^\nu=-1, \qquad f_{\mu\nu}n^\mu e^\nu_i=0.
$$
 If we decompose  over this basis another metric, for example,  $g_{\mu\nu}$, we obtain
\begin{eqnarray}
||g^{\mu\nu}||&=&
\left(
\begin{array}{cc}
-u^{-2}[n^\mu n^\nu] & u^ju^{-2}[n^\mu e^\nu_j]\\
u^iu^{-2}[e^\mu_i n^\nu] & (\gamma^{ij}-u^iu^ju^{-2})[e^\mu_i e^\nu_j]
\end{array}\right),\\
||g_{\mu\nu}||&=&\left(\begin{array}{cc}(-u^2+\gamma_{mn}u^m u^n)[n_\mu n_\nu] & -\gamma_{jk}u^k[n_\mu e_\nu^j]\\-\gamma_{ik}u^k[e_\mu^i n_\nu] & \gamma_{ij}[e_\mu^i e_\nu^j]\end{array}\right),
\end{eqnarray}
where new variables $u$, $u^i$ appear. Their role is, at least, three-fold: first, they appear in the relations between two pairs of lapse and shift functions:
\be
u=\frac{\bar N}{N},\qquad u^i=\frac{\bar N^i-N^i}{N},\label{eq:18}
\ee 
second, they are formed by projections of tensor $g^{\mu\nu}$ onto the basis $(n_\alpha,e_\alpha^i)$ constructed with the metric $f_{\mu\nu}$
\be
u=\frac{1}{\sqrt{-g^{\perp\perp}}}\equiv\frac{1}{\sqrt{-g^{\mu\nu}n_\mu n_\nu}},\qquad u^i=-\frac{g^{\perp i}}{g^{\perp\perp}}\equiv \frac{g^{\mu\nu}n_\mu e_\nu^i}{g^{\alpha\beta}n_\alpha n_\beta},
\ee
third, they are the coefficients that connect the two bases $(\bar n_\alpha,\bar e_\alpha^i)$ and $(n_\alpha,e_\alpha^i)$:
\be
\bar n_\mu=un_\mu,\quad \bar e^i_\mu=e^i_\mu-u^in_\mu,\quad \bar n^\mu=\frac{1}{u}n^\mu-\frac{u^i}{u}e^\mu_i,\quad \bar e^\mu_i=e^\mu_i.
\ee
As the bases $(n_\alpha,e_\alpha^i)$ and $(n^\alpha,e^\alpha_i)$ are formed with the help of metric $f_{\mu\nu}$, this metric has only 6 nontrivial components in decompositions over them
\begin{eqnarray}
||f^{\mu\nu}||&=&
\left(
\begin{array}{cc}
-[n^\mu n^\nu] & 0
\\
0
& \eta^{ij}[e^\mu_i e^\nu_j]
\end{array}\right),\\
||f_{\mu\nu}||&=&\left(\begin{array}{cc}-[n_\mu n_\nu] & 0
\\0
& \eta_{ij}[e_\mu^i e_\nu^j]\end{array}\right).
\end{eqnarray}

One can see that all the components of matrix $\mathsf{Y}^\mu_\nu=g^{\mu\alpha}f_{\alpha\nu}$ are functions of variables $u$, $u^i$, $\gamma_{ij}$ and $\eta_{ij}$, the same statement is valid for the invariants of this matrix, for example, $\mathrm{Tr}\left(\mathsf{Y}^n\right)$. Therefore the eigenvalues of $\mathsf{X}=\sqrt{\mathsf{Y}}$ are also dependent only on these variables.
In general case it is impossible to obtain the explicit expression for the dRGT potential in the form 
$$
U=N\tilde U(u,u^i,\eta_{ij},\gamma_{ij}).
$$
For this reason we  start with a general function $\tilde U$ and work with its formal derivatives
\begin{eqnarray}
 V&=&\frac{\partial \tilde U}{\partial u},\\
V_i&=&\frac{\partial \tilde U}{\partial u^i},
\end{eqnarray}
together with the function
$$
W=\tilde U-u\frac{\partial\tilde U}{\partial u}-u^i\frac{\partial\tilde U}{\partial u^i}.
$$
The Hamiltonian now appears in the following form
$$
{\rm H}=
\int d^3x 
\left[
N\left({\cal H}+
u\bar{\cal H}+
u^i\bar{\cal H}_i+{\cal H}_M+
\frac{2m^2}{\kappa}\tilde U \right)+ 
N^i\left({\cal H}_i+\bar{\cal H}_i +{\cal H}_{Mi}\right)\right].
$$
Then the derivatives of the total Hamiltonian over non-dynamical variables $N$, $N^i$, $u$, $u^i$ provide us with the primary relations:
\begin{eqnarray}
{\cal R}&=& \frac{\partial H}{\partial N}\equiv 
{{\cal H}}+u{\bar{\cal H}}+u^i{\bar{\cal H}}_i+{\cal H}_M+\frac{2m^2}{\kappa}\tilde U=0
,\\
{\cal R}_i&=&\frac{\partial H}{\partial N^i}\equiv {\cal H}_i+\bar{\cal H}_i+{\cal H}_{Mi},\\
 {\cal S}&=&  \frac{1}{N}\frac{\partial H}{\partial u}\equiv \bar{\cal H}+\frac{\partial {\cal H}_M}{\partial u}+\frac{2m^2}{\kappa}V=0,\\
{\cal S}_i&=&  \frac{1}{N}\frac{\partial H}{\partial u^i}\equiv \bar{\cal H}_i+\frac{\partial {\cal H}_M}{\partial u^i}+\frac{2m^2}{\kappa}V_i=0.
\end{eqnarray}
Then we can rewrite the Hamiltonian as follows
\be
{\rm H}=
\int d^3x 
\left[
N\left({\cal R}'+
u{\cal S}+
u^i{\cal S}_i \right)+ 
N^i{\cal R}_i\right],
\ee
where
\be
{\cal R}'= {\cal H}+\frac{2m^2}{\kappa}W,
\ee
For the potential $\tilde u$ of the general form equations 
\be
{\cal S}=0,\qquad {\cal S}_i=0,\label{eq:SS0}
\ee
may be solved for variables $u$, $u^i$, and then these solutions should be substituted into the expression of ${\cal R}'$ in order to make it a constraint, i.e. a function of canonical variables only. Ater that equations
\be
{\cal R}=0, \qquad {\cal R}_i=0,
\ee
occur 1st class constraints arising as a consequence of the space-time coordinate invariance of the bigravity. On-shell, i.e. on the surface of  variables satisfying Eqs.(\ref{eq:SS0}) they form the celebrated Poisson bracket algebra 
\begin{equation}
 \{ {\cal R}(x),{\cal R}(y)\}\approx\left(\eta^{ik}(x){\cal R}_k(x)+ \eta^{ik}(y){\cal R}_k(y)\right)\delta_{,i}(x,y),
\end{equation}
\begin{equation}
  \{ {\cal R}_i(x),{\cal R}_k(y)\}\approx{\cal R}_i(y)\delta_{,k}(x,y)+ {\cal R}_k(x)\delta_{,i}(x,y),
\end{equation}
\begin{equation}
  \{ {\cal R}_i(x),{\cal R}(y)\}\approx{\cal R}(x)\delta_{,i}(x,y),
\end{equation}
discussed by Dirac in the proposal to construct the field  quantization on the curved surfaces~\cite{Dirac}.
For the potential $\tilde U$ of the general form we obtain equations necessary to fulfil the above Poisson algebra
 \begin{eqnarray}
2\eta_{ik}\frac{\partial\tilde{U}}{\partial\eta_{jk}}+2\gamma_{ik}\frac{\partial\tilde{U}}{\partial\gamma_{jk}}-u^j\frac{\partial\tilde{U}}{\partial u^i}-\delta^j_i\tilde{U}&=&0,\label{eq:condition1} \\
 2u^j\gamma_{jk}\frac{\partial\tilde{U}}{\partial\gamma_{ik}}-u^i u\frac{\partial\tilde{U}}{\partial u}+\left(\eta^{ik}-u^2\gamma^{ik}-u^i u^k \right)\frac{\partial\tilde{U}}{\partial u^k}&=&0.\label{eq:condition2}
\end{eqnarray}
In order equation ${\cal S}=0$ to become a geniune constraint excluding the Boulware-Deser ghost degree of freedom it is necessary to put one more condition on the potential. This condition is the requirement that it should be  impossible to solve Eqs.(\ref{eq:SS0}) for auxiliary variables $u^a=(u,u^i)$, i.e. these equations are to be functionally dependent,
\be
\frac{D({\cal S},{\cal S}_i)}{D(u,u^i)}=
\det\left|\left|\frac{\partial^2}{\partial u^a\partial u^b}\left(\frac{2m^2}{\kappa}\tilde U+{\cal H}_M\right) \right|\right|\equiv\mbox{Hess}_{(4\times 4)}\left(\frac{2m^2}{\kappa}\tilde U+{\cal H}_M\right)=0.\label{eq:HessU}
\ee

Next we need to prove the existence of a secondary constraint $\Omega$, and the fact that it has a non-zero Poisson bracket with ${\cal S}$. For this purpose it is useful to apply a method suggested by Fairlie and Leznov~\cite{Leznov} on constructing implicit solutions of the homogeneous Monge-Amp\`ere equation. This trick has been first developed and applied to massive gravity by Comelli, Nesti, and Pilo~\cite{Comelli}. Combining 
Eqs.(\ref{eq:condition1}), (\ref{eq:condition2}) with the technique of Ref.\cite{Leznov} it is possible to prove that
$$
\{{\cal S}(x),{\cal S}(y)\}\approx 0,
$$
and therefore
$$
\dot{\cal S}=\{{\cal S},\mathrm{H}\}\approx\int d^3x N\{{\cal S},{\cal R}'\}=\int d^3x N\Omega =0\qquad \leftrightarrow \qquad \Omega=0.
$$
By means of the Jacobi identity it is also possible to show that in general
$$
\{\Omega,{\cal S}\}\ne  0,
$$ 
and so these two constraints are second class in the Dirac terminology. The auxiliary variable $u$ is fixed by the condition of preservation the secondary constraint in time
$$
\{\Omega,\mathrm{H}\}=\int d^3x N\left(\{\Omega,{\cal R}'\}+u\{\Omega,{\cal S}\}\right)=0.
$$ 
The other auxiliary variables $u^i$ are to be determined from equations ${\cal S}_i=0$. For this purpose it is necessary to require
\be
\det\left|\left|\frac{\partial{\cal S}_i}{\partial u^k}\right|\right|= \det\left|\left|\frac{\partial^2}{\partial u^i\partial u^k}\left(\frac{2m^2}{\kappa}\tilde U+{\cal H}_M\right) \right|\right|\equiv\mbox{Hess}_{(3\times 3)}\left(\frac{2m^2}{\kappa}\tilde U+{\cal H}_M\right)\ne 0,
\ee
and therefore the rank of the big $4\times 4$ Hessian (\ref{eq:HessU}) should be equal to three.
Finally, we come to the Hamiltonian formalism of bigravity with 12 pairs of canonical variables $(\gamma_{ij},\pi^{ij})$, $(\eta_{ij},\Pi^{ij})$, four 1st class constraints ${\cal R},{\cal R}_i$, and two 2nd class constraints ${\cal S},\Omega$. There are 7 gravitational degrees of freedom for such a gravitational potential.  

Here we followed articles~\cite{SolTch,SolTch2}. For the massive gravity case see also~\cite{Comelli}. 

\section{
Tetrad variables}
It is possible also to construct the Hamiltonian approach to bigravity with the tetrad variables connected with the metrics by formulas
\be
g^{\mu\nu}=E^\mu_AE^\nu_Bh^{AB},\qquad f_{\mu\nu}=F^A_\mu F_\nu^B h_{AB},\label{eq:tetrads}
\ee
where $h_{AB}$ is the  Minkowski metric
$$
h_{AB}=\mbox{diag}(-1,1,1,1),
$$
and $(E^\mu_A,E_\mu^A)$, $(F^A_\mu,F^\mu_A)$ are the mutually inverse tetrad matrices. 
In general the tetrad Hamiltonian formalism is more involved than the metric one. As 10 components of the metric tensor are replaced by 16 tetrad components the number of constraint equations increases. Therefore the constraint algebra becomes more extended.  In the same time the tetrad representation allows to obtain an explicit expression for the dRGT potential. Indeed, the matrix
\be
\mathsf{X}^\mu_\nu=E^{\mu A}F_{\nu A},\label{eq:matrixX}
\ee
occurs a square root of the matrix $\mathsf{Y}^\mu_\nu=g^{\mu\alpha}f_{\alpha\nu}$, 
if the symmetry equations 
\be
E^\mu_AF_\mu^B-E^{\mu B}F_{\mu A}=0, \label{eq:symmetry}
\ee
are valid. Then instead of Eqs.(\ref{eq:e_i}) it is suitable to apply  equivalent formulas for the symmetric polynomials 
\begin{eqnarray}
e_0&=&1,\nonumber\\
 e_1&=&\mathrm{Tr}\mathsf{X},\nonumber\\
e_2&=&\frac{1}{2}\left((\mathrm{Tr}\mathsf{X})^2-\mathrm{Tr}\mathsf{X}^2 \right),\nonumber\\
e_3&=&\frac{1}{6}\left((\mathrm{Tr}\mathsf{X})^3-3\mathrm{Tr}\mathsf{X}\mathrm{Tr}\mathsf{X}^2+2\mathrm{Tr}\mathsf{X}^3 \right),\nonumber\\
e_4&=&\frac{1}{24}\left( (\mathrm{Tr} \mathsf{X})^4 -6 (\mathrm{Tr} \mathsf{X})^2 \mathrm{Tr} \mathsf{X}^2 +3(\mathrm{Tr} \mathsf{X}^2)^2 +8 \mathrm{Tr} \mathsf{X} \mathrm{Tr} \mathsf{X}^3 -6 \mathrm{Tr} \mathsf{X}^4
\right)=\nonumber\\
&=&\det\mathsf{X}=\frac{\det ||F_{\mu a}|| }{\det ||E_{\mu a}|| }\equiv\frac{ \sqrt{-f} }{\sqrt{-g}}.\label{eq:traces}
\end{eqnarray}
It is convinient to start with the privileged form of a tetrad, and a bit later transform it to the general form. Starting first with the metric $g_{\mu\nu}$ we propose to use as such a simplified tetrad  the following set of four space-time vectors: the first vector $E^\alpha_0$ is the unit normal to the hypersurface of state $\bar n^\alpha$  constructed on the base of metric $g_{\mu\nu}$:
$$
g_{\mu\nu}\bar n^\mu e^\nu_i=0,\qquad g_{\mu\nu}\bar n^\mu\bar n^\nu=-1,
$$
the other three space-time vectors are the three tangential to the hypersurface vectors provided by the triads of the induced metric $\gamma_{ij}$
For this triads ${\bf e}_i^a$ we have equations similar to Eqs.(\ref{eq:tetrads}):
$$
\gamma_{ij}={\bf e}_i^a{\bf e}_j^b\delta_{ab},\qquad \delta_{ab}=\mbox{diag}(1,1,1),\qquad {\bf e}^i_a{\bf e}_{ib}=\delta_{ab}, \qquad {\bf e}^i_a{\bf e}_{ja}=\delta^i_j.
$$
Then the tangential space-time vectors are formed as follows
$$
E^\alpha_a={\bf e}^i_{a}\bar e^\alpha_i,\qquad \bar e^\alpha_i=e^\alpha_i\equiv\frac{\partial X^\alpha}{\partial x^i}.
$$
The inverse tetrads are 
$$
E_\mu^0=\bar n_\mu,\qquad E_\mu^a=\bar e_\mu^i{\bf e}_{ia},\qquad \bar e_\mu^i\equiv g_{\mu\nu}e^\nu_j\gamma^{ji}.
$$
Following this way for the metric $f_{\mu\nu}$ we also get a priviledged tetrad
\ber
{\cal F}_\mu^0&=&-n_\mu,\\
{\cal F}_\mu^a&=&e_\mu^j{\bf f}_{ja}, \qquad e_\mu^j=f_{\mu\nu}e^\nu_i\eta^{ij}.
\eer
The general tetrad $F^A_\nu$ arises as a result of a boost transformation 
  \begin{equation}
\Lambda^A_{ \ B}=\left(\begin{array}{cc} \varepsilon & p_b \\
p^a &{\cal P}^a_b \\ \end{array}
\right), \quad \varepsilon=\sqrt{1+p_ap^a},\quad {\cal P}^a_b =\delta^a_{ \ b}+\frac{1}{\varepsilon+1}p^a p_b\label{eq:Lorentz} \ ,
\end{equation}
of the priviledged one $F^A_\nu=\Lambda^A_{\ B}(p) {\cal F}^B_\nu$.
Here $$
p^a=p_a,\qquad
h_{AB}\Lambda^A_{ \ C}\Lambda^B_{ \ D}=h_{CD}.
$$
 Then we can derive the matrix $\mathsf{X}^\mu_\nu$ defined in (\ref{eq:matrixX}), and obtain the following
\begin{equation}
\mathsf{X}^\mu_\nu=\left(\begin{array}{cc} A[n^\mu n_\nu] & B^j[n^\mu e_{\nu j}] \\
C^i [e^\mu_i n_\nu] & D^{ij}[e^\mu_i e_{\nu j}] \\ \end{array}
\right), 
\end{equation}
where
\begin{eqnarray}
A&=&-\frac{\varepsilon}{u},\\
B^j&=&\frac{p_a{\bf f}^{ja}}{u},\\
C^i&=&\frac{u^i\varepsilon}{u}-p_a{\bf e}^{ia},\\
D^{ij}&=&-\frac{u^ip^a{\bf f}^{ja}}{u}+{\bf f}^{ja}{\cal P}_{ab}{\bf e}^{ib}.\label{eq:ABCD}
\end{eqnarray}
After straightforward calculation of all the symmetric polynomials (\ref{eq:traces}) 
we get an explicit expression for the dRGT potential which occurs linear in variables $u,u^i$
\be
\tilde U=uV+u^iV_i+W.
\ee
Functions $V$, $V_i$, and $W$ are given in Appendix. They depend on canonical variables ${\bf e}_{ia}$, ${\bf f}_{ia}$ and on auxiliary variable $p_a$.  Especially, $V_i$ occurs linear in $p_a$ 
\be
V_i=-p_a C_{ab}{\bf f}_{ib}.
\ee
As the potential term does not contain any velocities, it enters the Hamiltonian of bigravity without any changes. 

Next we are to find the corresponding Hamiltonian form for ${\cal L}_g$ and ${\cal L}_f$. This is familiar from GR~\cite{HamTetrads}. The new canonical  variables are triads ${\bf e}_{ia}$, ${\bf f}_{ia}$ together with their conjugate momenta $\pi^{ia}$, $\Pi^{ia}$, and their Poisson brackets are the following 
\begin{eqnarray}
\{{\bf f}_{ia}(x),\Pi^{jb}(y)\}&=&\{{\bf e}_{ia}(x),\pi^{jb}(y)\}=\delta_a^{b}\delta_i^{j}\delta(x,y),\\
\{{\bf f}_{ia}(x),{\bf e}_{jb}(y)\}&=&\{\pi^{ia}(x),\Pi^{jb}(y)\}=\{{\bf f}_{ia}(x),{\bf f}_{jb}(y)\}=0,\\
\{{\bf e}_{ia}(x),{\bf e}_{jb}(y)\}&=&\{\pi^{ia}(x),\pi^{jb}(y)\}=\{\Pi^{ia}(x),\Pi^{jb}(y)\}=0,
\end{eqnarray}
Momenta of the metric formalism are expressed through triad momenta as follows
\begin{eqnarray}
 \Pi^{ij}&=&\frac{1}{4}\left({\bf f}^{ia}\Pi^{ja}+{\bf f}^{ja}\Pi^{ia} \right),\label{eq:Pi}\\
 \pi^{ij}&=&\frac{1}{4}\left({\bf e}^{ia}\pi^{ja}+{\bf e}^{ja}\pi^{ia} \right).\label{eq:pi}
\end{eqnarray}
Then the Poisson brackets of the metric momenta are nonzero off-shell 
\begin{eqnarray}
 \{\Pi^{ij}(x),\Pi^{k\ell}(y)\}&=&\frac{1}{4}\left(\eta^{ik}{\cal M}^{j\ell}+\eta^{i\ell}{\cal M}^{jk}+\eta^{jk}
{\cal M}^{i\ell}+\eta^{j\ell}{\cal M}^{ik} \right),\\
\{\pi^{ij}(x),\pi^{k\ell}(y)\}&=& \frac{1}{4}\left(\gamma^{ik}\bar{\cal M}^{j\ell}+\gamma^{i\ell}\bar{\cal M}^{jk}+\gamma^{jk}\bar{\cal M}^{i\ell}+\gamma^{j\ell}\bar{\cal M}^{ik} \right).
\end{eqnarray}
Here the new constraints specific for the tetrad approach appear 
\begin{eqnarray}
 {\cal M}^{ij}&=&\frac{1}{4}L_{ab}{\bf f}^{ja}{\bf f}^{ib}\equiv\frac{1}{4}\left({\bf f}^{ia}\Pi^{j}_a-{\bf f}^{ja}\Pi^i_a \right)=0, \\
\bar{\cal M}^{ij}&=&\frac{1}{4}\bar L_{ab}{\bf e}^{ja}{\bf e}^{ib}\equiv\frac{1}{4}\left({\bf e}^{ia}\pi^{j}_a-{\bf e}^{ja}\pi^i_a \right)=0,
\end{eqnarray}
or in a different form,
\begin{eqnarray}
 L_{ab}&=&{\bf f}_{ia}\Pi^i_b-{\bf f}_{ib}\Pi^i_a=0,\label{eq:Labf}\\
 \bar L_{ab}&=&{\bf e}_{ia}\pi^i_b-{\bf e}_{ib}\pi^i_a=0.\label{eq:Labg}
\end{eqnarray}
The combined Hamiltonian, of course, contain these constraints
\ber
\mathrm{H}_{g+f}&=&\mathrm{H}_g+\mathrm{H}_f=\nonumber\\
&=& \int d^3x\left(\bar N\bar{\cal H}+\bar N^i\bar{\cal H}_i+\bar\lambda^{ab}\bar L_{ab} \right)+\nonumber\\
&+&  \int d^3x\left( N{\cal H}+ N^i{\cal H}_i+\lambda^{ab} L_{ab} \right),
\eer
together with the constraints $\bar{\cal H}$, $\bar{\cal H}_i$, ${\cal H}$, ${\cal H}_i$ known from the metric Hamiltonian (\ref{eq:constraint1}) -- (\ref{eq:constraint02}). As in bigravity the two metrics are mixed in the potential, only diagonal rotations of triads ${\bf e}_{ia}$, ${\bf f}_{ia}$ leave the Hamiltonian invariant. Therefore symmetric combinations
\be
L^+_{ab}=\bar L_{ab}+L_{ab}=0,\label{eq:Lplus}
\ee
become 1st class constraints, whereas constraints
\be
L^-_{ab}=\bar L_{ab}-L_{ab}=0,\label{eq:Lminus}
\ee
are 2nd class.

Finally, we are to take into account the symmetry conditions (\ref{eq:symmetry}). In the Hamiltonian variables they take the following form
\begin{eqnarray}
G_a&\equiv&p_a+up_b{\bf  f}_j^b{\bf  e}^j_a-u^j{\cal P}_{ab}{\bf  f}_j^b=0,\label{eq:const1}\\
G_{ab}&\equiv&{\bf f}_{ic}{\cal P}_{c[a}{\bf e}^i_{b]}=0\label{eq:const2}.
\end{eqnarray}
The Hamiltonian of bigravity now appears in the following form
\begin{eqnarray}
{\rm H}&=&\int d^3x N\left[
{\cal H}+u\left(\bar{\cal H}+\frac{2m^2}{\kappa}V\right)+u^i\left(\bar{\cal H}_i+\frac{2m^2}{\kappa}V_i\right)+\frac{2m^2}{\kappa}W\right]+\nonumber\\
&+&\int d^3x\left[N^i({\cal H}_i+\bar{\cal H}_i)+\lambda^+_{ab}L^+_{ab}+ \lambda^-_{ab}\bar L^-_{ab}+\Lambda^aG_a+\Lambda^{ab}G_{ab}\right].\label{eq:H_0}
\end{eqnarray}
When two types of matter, f-matter and g-matter, separately and minimally couple to metrics $f_{\mu\nu}$ and $g_{\mu\nu}$, their contributions can be incorporated into  constraints   $\bar{\cal H}$, $\bar{\cal H}_i$, ${\cal H}$, ${\cal H}_i$. 
Besides the canonical variables the Hamiltonian depend on the Lagrangian multipliers $\lambda^+_{ab}$, $\lambda^-_{ab}$, $\Lambda^a$, $\Lambda^{ab}$, providing Eqs. (\ref{eq:Lplus})--(\ref{eq:const2}) and on other auxiliary variables. Varying over  $u$, $u^i$, $N$, $N^i$ we obtain the following equations
\ber
{\cal S}&=& \bar{\cal H}+\frac{2m^2}{\kappa}V=0,\\
{\cal S}_i&=& \bar{\cal H}_i+\frac{2m^2}{\kappa}V_i,\label{eq:Si}\\
{\cal R}&=&{\cal R}'+u{\cal S}+u^i{\cal S}_i=0,\quad \mbox{where} \quad {\cal R}'={\cal H}+\frac{2m^2}{\kappa}W,\\
{\cal R}_i&=&{\cal H}_i+\bar{\cal H}_i=0.
\eer
We do not vary over variable $p_a$ that enter (\ref{eq:H_0}) in non-linear way.
Eqs.(\ref{eq:const1}) can be solved for $u^i$
\be
u^i={\bf f}^{ib}\left(\frac{p_b}{\varepsilon}+up_a{\bf f}_j^a{\bf e}^{jc}{\cal P}^{-1}_{cb} \right),
\ee
where
\be
{\cal P}^{-1}_{cb}=\delta_{cb}-\frac{p_c p_b}{\varepsilon(\varepsilon+1)}.
\ee
Eqs. (\ref{eq:Si}) contain variable $p_a$ in a linear way and so can be easily solved for it
\be 
p_a=\frac{2\kappa}{m^2 }||C_{ab}||^{-1}f^{ib}\bar{\cal H}_i.
\ee 
Eqs.(\ref{eq:const2}) depend on the canonical variables ${\bf e}^i_b$, ${\bf f}_{ic}$, and on the auxiliary variable $p_a$. After the exclusion of $p_a$ $G_{ab}$ become constraints. They have nonzero Poisson brackets with $L^-_{ab}$ and together form 6 constraints of the 2nd class.

In the tetrad Hamiltonian formalism there are 18 pairs of the canonical variables $({\bf e}_{ia},\pi^{ia})$, $({\bf f}_{ia},\Pi^{ia})$, 7 constraints of the 1st class ${\cal R}'$, ${\cal R}_i$, $L^+_{ab}$ and 8 constraints of the 2nd class ${\cal S}$, $\Omega$, $L^-_{ab}$, $G_{ab}$.
Of course, we again have 7 gravitational degrees of freedom. The auxiliary variables $p_a$ are found from equations ${\cal S}_i=0$ which are linear in $p_a$. Variables $u^i$ are found from equations $G_a=0$ linear in $u^i$. The last auxiliary variable $u$ can be found from
\be
\dot\Omega=0=\int d^3x N\left(\{\Omega,{\cal R}'\}+u\{\Omega,{\cal S}\} \right),
\ee 
this is also a linear equation.

The above exposition followed articles~\cite{Sol,Soloviev:2014eea}.
Different treatments can be found in Refs.~\cite{Kluson_tetrad,Krasnov,Alexandrov}.

\section{
Minisuperspace}
Let us look at the cosmological problem, i.e. consider the Hamiltonian evolution of a homogeneous and isotropic Universe in the bigravity theory. For simplicity and brevity only the spatially flat case will be treated. Most of attention will be paid to the theory with the dRGT potential.

In general, the bigravity Hamiltonian is a sum of kinetic and potential gravitational terms, composed of two metrics $g_{\mu\nu}$ and $f_{\mu\nu}$, plus one, or two matter contributions.

We take the Friedmann-Robertson-Walker ansatze for the two metrics
\ber
f_{\mu\nu}&=&(-N^2(t),\omega^2(t)\delta_{ij}),\qquad \sqrt{-f}=N\omega^3,\label{eq:f}\\
g_{\mu\nu}&=&(-N^2(t)u^2(t),\xi^2(t)\delta_{ij}),\qquad  \sqrt{-g}=Nu\xi^3,\label{eq:g}
\eer
and introduce a new variable $r=\omega/\xi$.

\subsection{Gravitational potential terms}
For the admitted metric tensors of the form (\ref{eq:f}), (\ref{eq:g}) we obtain
\be
g^{-1}f=g^{\mu\alpha}f_{\alpha\nu}=\mathrm{diag}\left(u^{-2},
r^2\delta_{ij}\right).\label{eq:Xmini}
\ee
Let us suppose
\be
N>0,\quad u>0,\quad \xi>0,\quad \omega>0.
\ee
In fact, in calculating the matrix square root of the positive diagonal matrix (\ref{eq:Xmini}) we consider only one (positive) square root for each positive expression~\footnote{In this special case nothing serious will change if we take another sign for some eigenvalues.}
\be
\mathsf{X}=\sqrt{g^{-1}f}=\mathrm{diag}\left(+\sqrt{u^{-2}},
+\sqrt{r^2}\delta_{ij}\right)\equiv\mathrm{diag}\left(u^{-1},
r\delta_{ij}\right),
\ee
and eigenvalues of this matrix are determined from the following equation
\be
\det(\mathsf{X}-\lambda I)=0.
\ee
It is easy to find that
\be
 \lambda_1=u^{-1},\quad \lambda_2=\lambda_3=\lambda_4=
r,
\ee
and then construct the symmetric polynomials:
\ber
e_0&=&1\\
e_1&=&\lambda_1+\lambda_2+\lambda_3+\lambda_4=\frac{1}{u}+3r,\\
e_2&=&\lambda_1\lambda_2+\lambda_1\lambda_3+\lambda_1\lambda_4+\lambda_2\lambda_3+\lambda_2\lambda_4+\lambda_3\lambda_4=\frac{3r}{u}+3r^2 ,\\
e_3&=&\lambda_1\lambda_2\lambda_3+\lambda_2\lambda_3\lambda_4+\lambda_1\lambda_3\lambda_4+\lambda_1\lambda_2\lambda_4=\frac{3r^2}{u}+r^3,\\
e_4&=&\lambda_1\lambda_2\lambda_3\lambda_4=\frac{r^3}{u}.
\eer
The dRGT potential is a linear combination of these polynomials which occurs a linear function of variable $u$:
\be
 U=\sqrt{-g}\sum_{i=0}^4\beta_i e_i=Nu\xi^3\sum_{i=0}^4\beta_i e_i=N(uV+W),
\ee
where
\ber
V&=&\beta_0{\xi}^3+3\beta_1{\xi}^2\omega+3\beta_2{\xi}\omega^2+\beta_3\omega^3\equiv \xi^3B_0(r),\\
W&=&  \beta_1{\xi}^3+3\beta_2{\xi}^2\omega+3\beta_3{\xi}\omega^2+\beta_4\omega^3\equiv \xi^3B_1(r),\\
B_i(r)&=&\beta_i+3\beta_{i+1}r+3\beta_{i+2}r^2+\beta_{i+3}r^3.
\eer

\subsection{Gravitational kinetic terms}
If we start with the Hilbert-Einstein gravitational Lagrangian density for the space-time metric ${\cal G}_{\mu\nu}$
\be
{\cal L}_{\cal G}=\frac{1}{\kappa_{\cal G}}\sqrt{-{\cal G}}{\cal G}^{\mu\nu}R_{\mu\nu},
\ee
and exploit the cosmological ansatze
\be
{\cal G}_{\mu\nu}=(-{\cal N}^2(t),a^2(t)\delta_{ij}),\label{eq:calGmini}
\ee
we get
\be
{\cal L}_{\cal G}={\cal T}_{\cal G}=-\frac{6a^3{\cal N}}{\kappa} \left(\frac{\dot a}{{\cal N}a}\right)^2\equiv -\frac{6a^3{\cal N}}{\kappa}  H^2,
\ee
where $H$ is the Hubble constant.

Define the momentum canonically conjugate to $a$ 
\be
\pi_a=\frac{\partial {\cal L}_G}{\partial \dot a}=-\frac{12 a}{{\cal N}\kappa}\dot a=-\frac{12 a^2 }{\kappa} H,\qquad H=-\frac{\kappa\pi_a}{12 a^2},\qquad \{a,\pi_a\}=1.
\ee
It is suitable to use $H$ instead of $\pi_a$, then we will use non-canonical Poisson bracket:
\be
\{a,H\}=-\frac{\kappa}{12 a^2}.
\ee
In bigravity we get the kinetic part of the Hamiltonian in the following form
\be
 \mathrm{H}_{kinetic}= {\cal T}_f+{\cal T}_g=
-\frac{6\omega^3{ N}}{\kappa_f}  H_f^2 -\frac{6\xi^3{ Nu}}{\kappa_g}  H_g^2.
\ee
This part generates the kinematical Hamiltonian equations
\be
\dot\omega=\{\omega,
 \mathrm{H}_{kinetic}\}=N\omega H_f,\qquad  \dot\xi=\{\xi,
 \mathrm{H}_{kinetic}\}=Nu\xi H_g, \label{eq:kinematical}
\ee
equivalent to the definition of two Hubble constants $H_f$, $H_g$. It is easy to derive an equation for the evolution of the relative scale factor  $r$
\be
\dot r=Nr(H_f-uH_g).\label{eq:rdot}
\ee

\subsection{Matter terms}
As an example of matter, we take the scalar field having a minimal interaction with gravity
\be
{\cal L}_\phi=\sqrt{-{\cal G}}\left(-\frac{1}{2}{\cal G}^{\mu\nu}\phi_{,\mu}\phi_{,\nu}-{\cal U}(\phi) \right),
\ee
where ${\cal G}_{\mu\nu}$ is the relevant metric of the form given by Eq.(\ref{eq:calGmini}) (there will be different cases below), where $\phi_{,\mu}={\partial\phi}/{\partial x^\mu}$. In the homogeneous cosmology $\phi=\phi(t)$ and
\be
{\cal L}_\phi={\cal T}_\phi[\dot \phi]-{\cal U}_\phi,\qquad {\cal H}_\phi={\cal T}_\phi[\pi_\phi]+{\cal U}_\phi.
\ee
Scalar fields may be of special interest in studying inflation in the bigravity~\cite{Sakakihara2015}.
If this matter field interacts with only one metric ${\cal G}_{\mu\nu}$, there is a straightforward correspondence between the velocity and the momentum
\be
{\cal T}_\phi=\frac{a^3{\dot\phi}^2}{2{\cal N}},\qquad \pi_\phi=\frac{\partial{\cal T}_\phi}{\partial \dot\phi}=\frac{a^3}{{\cal N}}\dot\phi,\qquad \dot\phi=\frac{{\cal N}}{a^3}\pi_\phi.
\ee
We can also introduce density $\rho$ and pressure $p$ of matter according to the formulas
\be
{\cal H}_\phi={\cal N}a^3\rho, \qquad \rho=\frac{\pi_\phi^2}{2a^6}+{\cal U}(\phi), \qquad p=\frac{\pi_\phi^2}{2a^6}-{\cal U}(\phi).
\ee
Below we will also consider the case when a scalar field simultaneously couples to metrics $g_{\mu\nu}$ and $f_{\mu\nu}$. Then
\be
{\cal T}_\phi=\frac{\xi^3{\dot\phi}^2}{2{\bar N}}+\frac{\omega^3{\dot\phi}^2}{2{N}},\qquad \pi_\phi=\frac{\partial{\cal T}_\phi}{\partial \dot\phi}=\left(\frac{\xi^3}{u}+\omega^3\right)\frac{\dot\phi}{N},\qquad \dot\phi=\frac{N\pi_\phi}{\frac{\xi^3}{u}+\omega^3},
\ee
and therefore
\be
{\cal T}_\phi=\frac{N\pi^2_\phi}{2\left(\frac{\xi^3}{u}+\omega^3\right)},\qquad {\cal U}_\phi=N\left(u\xi^3+\omega^3\right){\cal U},
\ee
\be
 {\cal H}_\phi=N(u\xi^3+\omega^3)\left(\frac{\pi^2_\phi}{2\left(\frac{\xi^3}{u}+\omega^3\right)(u\xi^3+\omega^3)}+{\cal U}(\phi) \right). 
\ee
Here ${\cal H}_\phi$ is a nonlinear function of $u$.

\subsection{List of GR formulas given for reference}
\be
{\cal L}_g=\frac{1}{\kappa_g}\sqrt{-g}(g^{\mu\nu}R_{\mu\nu}-2\Lambda)+{\cal L}_M(\phi^A, g_{\mu\nu}),\qquad \kappa_g=16\pi G,
\ee
\be
{\cal L}_\phi=\sqrt{-g}\left(-\frac{1}{2}g^{\mu\nu}\phi_{,\mu}\phi_{,\nu}-{\cal U}(\phi) \right),
\ee
\be
R^{\mu\nu}-\frac{1}{2}g^{\mu\nu}(R-2\Lambda)=\frac{\kappa_g}{2}T^{\mu\nu}, \qquad T^{\mu\nu}=\frac{2}{\sqrt{-g}}\frac{\partial{\cal L}_M}{\partial g_{\mu\nu}}.
\ee
\be
g_{\mu\nu}=(-{\bar N}^2(t),\xi^2(t)\delta_{ij}),\qquad  \sqrt{-g}={\bar N}\xi^3,\qquad \phi=\phi(t).
\ee
\be
{\cal L}_g={\bar N}\xi^3\left(-\frac{6}{\kappa_g}\left(\frac{\dot \xi}{{\bar N}\xi}\right)^2
-2\Lambda\right)
, \qquad H_g=\frac{\dot \xi}{{\bar N}\xi}, \qquad \pi_\xi=-\frac{12\xi^2}{\kappa_g}H_g.
\ee
\be
{\cal L}_\phi={\bar N}\xi^3\left(\frac{1}{2}\left(\frac{{\dot\phi}}{{\bar N}}\right)^2-{\cal U}(\phi)\right), \qquad \pi_\phi=
\frac{\xi^3}{{\bar N}}\dot\phi.
\ee
\be
\rho=\frac{{\pi_\phi}^2}{2\xi^6}+U(\phi)=\frac{1}{2}\left(\frac{{\dot\phi}}{{\bar N}}\right)^2-{\cal U}(\phi),\qquad p=\frac{{\pi_\phi}^2}{2\xi^6}-{\cal U}(\phi).
\ee
\be
\mathrm{H}={\bar N}\xi^3\left(
-\frac{6H_g^2}{\kappa_g}+\rho+\frac{2\Lambda}{\kappa_g}\right), \qquad 
\{ \xi,H_g \}
=-\frac{\kappa_g}{12\xi^2}.
\ee
\be
-\frac{6H_g^2}{\kappa_g}+\rho+\frac{2\Lambda}{\kappa_g}=0,\qquad\leftrightarrow\qquad \left(\frac{\dot \xi}{{\bar N}\xi}\right)^2=\frac{8\pi G}{3}\rho+\frac{\Lambda}{3}.\label{eq:Friedmann}
\ee
\ber
\dot \xi&=&\bar N\xi H_g,\\
\dot H_g&=&-\frac{{\bar N}\kappa_g}{4}(\rho+p)
=-4\pi G{\bar N}(\rho +p),\label{eq:GRT2}\\
\dot\rho&=&-3\frac{\dot \xi}{\xi}(\rho+p).
\eer
If $\Lambda$ and the matter equation of state $p=p(\rho)$ are known, the initial condition can be taken for $\rho$ or for $H_g$. $\bar N$ is arbitrary in accordance to the freedom of  time reparametrization.  

\section{Matter couplings}
\subsection{f-matter and g-matter}
The first works on cosmology of bigravity~\cite{Volkov:2011an,vonStrauss:2011mq,Comelli:2011zm,Akrami:2012vf} appeared rather soon after the proposal of the bigravity theory with the dRGT potential made by Hassan and Rosen~\cite{HR_bi}. In trying to solve the dark energy problem it is enough to admit g-matter only, but if also dark matter is under study then f-matter is also required~\cite{DM}.

If there are two species of matter: f-matter and g-matter, and each one has a minimal coupling to $f_{\mu\nu}$ or $g_{\mu\nu}$, then 
\be
\mathrm{H}=\mathrm{H}_{potential}+\mathrm{H}_{kinetic}+\mathrm{H}_{matter},
\ee
where
\ber
\mathrm{H}_{potential}&=&\frac{2m^2}{\kappa}N(uV+W),\\
\mathrm{H}_{kinetic}&=&\frac{1}{\kappa_f}N\left(-6\omega^3 H_f^2 \right)+\frac{1}{\kappa_g}Nu\left(-6\xi^3 H_g^2 \right),\\
\mathrm{H}_{matter}&=&N\omega^3\rho_f+Nu\xi^3\rho_g.
\eer
The Hamiltonian can also be written as follows
\be
\mathrm{H}=N{\cal R}\equiv N{\cal R}'+Nu{\cal S},
\ee
where primary constraints are ${\cal R}\equiv {\cal R}'+u{\cal S}$ and ${\cal S}$, or better, ${\cal R}'$ and ${\cal S}$: 
\ber
{\cal R}'&=& -\frac{6}{\kappa_f}\omega^3H_f^2+\omega^3\rho_f+\frac{2m^2}{\kappa}\xi^3B_1(r)=0\label{eq:R1},\\
{\cal S}&=& -\frac{6}{\kappa_g}\xi^3H_g^2+\xi^3\rho_g+\frac{2m^2}{\kappa}{\xi}^3B_0(r)=0.\label{eq:S1}
\eer
For the scalar sources we have
\ber
\rho_f&=&\frac{\Pi_\Phi^2}{2\omega^6}+ {\cal U}_f(\Phi),\\
\rho_g&=&\frac{\pi_\phi^2}{2\xi^6}+ {\cal U}_g(\phi).
\eer
The constraints (\ref{eq:R1}), (\ref{eq:S1}) may be written as a couple of Friedmann equations:
\ber
H_f^2&=&\frac{\kappa_f}{6}\rho_f+\frac{\Lambda_f}{3},\qquad \Lambda_f=m^2\frac{\kappa_f}{\kappa}\frac{B_1(r)}{r^3},\label{eq:Hf}\\
H_g^2&=&\frac{\kappa_g}{6}\rho_g+\frac{\Lambda_g}{3},\qquad \Lambda_g=m^2\frac{\kappa_g}{\kappa}B_0(r).\label{eq:Hg}
\eer 
We see that, in contrast to GR (\ref{eq:Friedmann}), the cosmological terms naturally appear in the bigravity, and they are not constants, but dynamical quantities.
Hamiltonian equations for the matter fields provide us with a couple of conservation laws
\ber
\dot\rho_f+3\frac{\dot\omega}{\omega}(\rho_f+p_f)&=&0,\\
\dot\rho_g+3\frac{\dot\xi}{\xi}(\rho_g+p_g)&=&0.
\eer

The secondary constraint arises in order to preserve the primary constraint during the evolution
\be
\dot{\cal S}=\{{\cal S},\mathrm{H}\}=N\{{\cal S},{\cal R}'\}\equiv N\Omega=0,
\ee
where 
\be
\Omega\equiv\{{\cal S},{\cal R}'\}=\frac{6m^2}{\kappa}\left(\omega H_f-\xi H_g \right)\left(\beta_1 \xi^2+2\beta_2 \xi\omega+\beta_3\omega^2\right)=0.\label{eq:Omega}
\ee
As the secondary constraint is factorized 
\be
\Omega=\Omega_1\Omega_2,\label{eq:factor}
\ee
 there are two branches of solutions for it:
\be
\Omega_1=0,\quad \leftrightarrow\quad H_g=r H_f, \label{eq:1st}
\ee
and
\be
\Omega_2=0,\quad \leftrightarrow\quad\beta_1 +2\beta_2 r+\beta_3r^2=0.\label{eq:2nd}
\ee
Below we consider both two
cases.

Variable $u$ is fixed by the following requirement
\be
\dot\Omega=\{\Omega,\mathrm{H}\}=N\left(\{\Omega,{\cal R}'\}+u\{\Omega,{\cal S}\} \right)=0,
\ee
here we also have two cases:
\be
u=-\frac{\{\Omega_1,{\cal R}'\}}{\{\Omega_1,{\cal S}\}},\qquad \mbox{or} \qquad  u=-\frac{\{\Omega_2,{\cal R}'\}}{\{\Omega_2,{\cal S}\}} .\label{eq:48}
\ee
For the two Hubble constants we get dynamical Hamiltonian equations
\ber
\dot H_f&=&\{H_f,\mathrm{H}\}=\{H_f,\omega\}\frac{\partial\mathrm{H}}{\partial\omega},\\
\dot H_g&=&\{H_g,\mathrm{H}\}=\{H_g,\xi\}\frac{\partial\mathrm{H}}{\partial\xi},
\eer
which appear in the following explicit form (with account for Eqs.(\ref{eq:R1}), (\ref{eq:S1}))
\ber
\dot H_f &=&-\frac{N\kappa_f}{4}\left[\rho_f+p_f+\left(1-ur \right)\frac{2m^2}{\kappa}\frac{D_1(r)}{r^3}  \right],\label{eq:Hfdot}\\
\dot H_g &=&-\frac{Nu\kappa_g}{4}\left[\rho_g+p_g-\left(1-ur\right)\frac{2m^2}{\kappa}\frac{D_1(r)}{u} \right]\label{eq:Hgdot},
\eer
where
\be
D_i(r)=\beta_i+2\beta_{i+1}r+\beta_{i+2}r^2.
\ee

\subsubsection{Massive gravity on Minkowskian background}

If we suppose $f_{\mu\nu}$ to be a non-dynamical  Minkowskian metric
\be
N=1,\qquad \omega=1,\qquad r=\frac{1}{\xi},
\ee
 and, of course, exclude f-matter 
\be
\rho_f=0,\qquad p_f=0,
\ee
then we get a massive gravity theory
\ber
\mathrm{H}&=&u{\cal S}+\frac{2m^2}{\kappa}W, \\
{\cal S}&=&{\xi}^3\left[-\frac{6}{\kappa_g}H_g^2+\rho_g+\frac{2m^2}{\kappa}B_0\left(\frac{1}{\xi}\right)\right]=0,\\
\Omega&=&-\frac{12m^2}{\kappa}\xi H_g \left(\beta_1 \xi^2+2\beta_2 \xi+\beta_3\right)=0.
\eer
From the last equation it follows that there are no non-static homogeneous and isotropic spatially flat cosmological solutions. This result has been obtained  in Refs.~\cite{PhysRevD.84.124046,Gumrukcuoglu:2011ew}.

\subsubsection{First branch of  bigravity}
From Eq.(\ref{eq:1st}) we get a relation between the two Hubble constants and taking into account the Friedmann equations (\ref{eq:Hf}), (\ref{eq:Hg})  we can derive the equation relating the two matter energy densities
\ber
H_f&=&r^{-1}H_g,\label{eq:HgHf}\\
\rho_g&=&\mu r^2\rho_f+\frac{2m^2}{\kappa}\left[\mu \frac{B_1(r)}{r}-B_0(r)\right],\label{eq:rhorho}
\eer
where
\be
\mu=\frac{\kappa_f}{\kappa_g}.
\ee
The first of  Eqs.(\ref{eq:48}) gives
\be
u=\frac{\mu r^3(\rho_f+3p_f)+\frac{2m^2}{\kappa}\left(\mu (B_1-3rD_2)+3r^2D_1
 \right)}
{r\left[r(\rho_g+3p_g)+\frac{2m^2}{\kappa}\left(
r(B_0-3D_0)      +3\mu D_1   
\right)\right]}.
\ee
Eq. (\ref{eq:rdot}) aquires the following form
\be
\dot r=(1-ru)NH_g,
\ee
and a fixed point for $r$ appears at
\be
r=\frac{1}{u}.
\ee 
It corresponds to proportional space-time metrics $f_{\mu\nu}=r^2g_{\mu\nu}$. 

In general case for the expression $1-ur$ we have the following formula
\be
1-ur=\frac{(\rho_g+3p_g)-\mu r^2(\rho_f+3p_f)+\frac{4m^2}{\kappa}\left(\frac{\mu B_1}{r}-B_0 \right)}
{(\rho_g+3p_g)+\frac{2m^2}{\kappa}\left(B_0-3D_0 +\frac{3\mu D_1}{r}\right)}.\label{eq:1ur}
\ee

In order to solve the system of Hamiltonian equations one can take as the initial data
$\rho_g(t_0)$ and $H_g(t_0)$. Then the constraint ${\cal S}$
\be
H_g^2=\frac{\kappa_g}{6}\rho_g+\frac{\Lambda_g}{3},
\ee
allows to find $\Lambda_g(t_0)$, and next, $r(t_0)=r_0$ can be found as one of the roots for the cubic equation
\be
B_0(r_0)=\frac{\Lambda_g(t_0)}{m^2}\frac{\kappa}{\kappa_g}.
\ee
Then constraint $\Omega_1$: 
\be
H_f=r^{-1}H_g,
\ee
gives us $H_f$. The constraint ${\cal R}'$:
\be
H_f^2=\frac{\kappa_f}{6}\rho_f+\frac{\Lambda_f}{3}, \qquad \Lambda_f=\mu m^2\frac{\kappa_g}{\kappa}\frac{B_1(r)}{r^3},
\ee
provides us with $\rho_f$. With account for Eq.(\ref{eq:1ur}) and equations of state $p_g=p_g(\rho)$, $p_f=p_f(\rho)$
it is possible to integrate the dynamical equations
\ber
\dot r&=&(1-ru)NH_g,\nonumber\\
\dot\rho_g&=&-3NuH_g(\rho_g+p_g),\nonumber\\
\dot\rho_f&=&-3NH_f(\rho_f+p_f),\nonumber\\
\dot H_g &=&-\frac{N\kappa_g}{4}\left[u(\rho_g+p_g)-\left(1-ur\right)\frac{2m^2}{\kappa}D_1(r) \right],\nonumber\\
\dot H_f &=&-\frac{N\kappa_f}{4}\left[\rho_f+p_f+\left(1-ur \right)\frac{2m^2}{\kappa}\frac{D_1(r)}{r^3}  \right],\nonumber
\eer
where $N(t)$ is an arbitrary monotonic function responsible for a freedom of the time reparametrization.

Besides the general case, when $1-ur$ is given by Eq.(\ref{eq:1ur}), 
it is tempting to consider the case when both types of matter have the same equation of state
\be
p_g=w\rho_g,\qquad p_f=w\rho_f,
\ee
then Eq.(\ref{eq:1ur}) is simplified, and is looking as follows 
\be
1-ur=\frac{3(\frac12+w)\left(\mu\frac{B_1}{r}-B_0 \right)
}{B_0-3D_0+\frac{3\mu D_1}{r}+\frac{\kappa}{2m^2} (1+3w)\rho_g
}.\label{eq:1ursimple}
\ee

\subsubsection{Second branch of bigravity}
In the second case (\ref{eq:2nd})  the value of variable $r$ is a constant:
\be
D_1(r)\equiv\beta_1 +2\beta_2 r+\beta_3r^2=0,\qquad \leftrightarrow \qquad 
r=\frac{-\beta_2\pm\sqrt{\beta_2^2-\beta_1\beta_3}}{\beta_3}.\label{eq:rfixed}
\ee
Then from the kinematical equation (\ref{eq:rdot}) it follows
\be
u=\frac{H_f}{H_g},
\ee
The last equation is not a constraint, in contrast to (\ref{eq:HgHf}), therefore the Friedmann equations (\ref{eq:Hf}), (\ref{eq:Hg}), and correspondingly, the dynamics for metrics $g_{\mu\nu}$, $f_{\mu\nu}$  decouple in this case. Dynamics for the two metrics are the same as in GR with the cosmological constants
\be
\Lambda_g=m^2\frac{\kappa_g}{\kappa}B_0(r), \qquad \Lambda_f= \mu m^2\frac{\kappa_g}{\kappa}\frac{B_1(r)}{r^3}.
\ee
Eqs.(\ref{eq:Hfdot}), (\ref{eq:Hgdot}) are 
as follows
\ber
\dot H_f &=&-\frac{N\kappa_f}{4}\left(\rho_f+p_f\right),\\
\dot H_g &=&-\frac{Nu\kappa_g}{4}\left(\rho_g+p_g\right),
\eer
and are the same as in GR (\ref{eq:GRT2}). If we put ourselves into g-world, then the only artifact of the bigravity will be the fixed value of $\Lambda_g$.

The full set of equations consists of the algebraic ones
\ber
r&=&\frac{-\beta_2\pm\sqrt{\beta_2^2-\beta_1\beta_3}}{\beta_3},\\
H_g^2&=&\frac{\kappa_g}{6}\rho_g+\frac{\Lambda_g}{3},\\
H_f^2&=&\frac{\kappa_f}{6}\rho_f+\frac{\Lambda_f}{3},\\
u&=&\frac{H_f}{H_g},
\eer
and the dynamical ones
\ber
\dot H_g &=&-\frac{Nu\kappa_g}{4}\left(\rho_g+p_g\right),\\
\dot H_f &=&-\frac{N\kappa_f}{4}\left(\rho_f+p_f\right),\\
\frac{\dot \xi}{\xi}&=&\frac{\dot\omega}{\omega}=-\frac{\dot\rho_g}{3(\rho_g+p_g)}=-\frac{\dot\rho_f}{3(\rho_f+p_f)}= NH_f.
\eer

\subsection{g-matter without f-matter}
To obtain  non-trivial results for the cosmology of bigravity which do not contradict the experimental evidence it is enough to study the case when f-matter is absent first considered in articles~\cite{Volkov:2011an,vonStrauss:2011mq,Comelli:2011zm}. In general, a solution allows for an evolution starting from a matter dominated Universe and evolving to de Sitter geometry at late times. As there are many parameters $\beta_i$, $i=0,\ldots,4$, and $\kappa_f$, the possible scenaria are diverse, including the cyclic Universe without the cosmological singularity. Even the elaborated statistical study presented, for example, in article~\cite{Akrami:2012vf} does not alow to fix the values of free parameters. 

Let us suppose $\rho_f=0=p_f$. This case is  discussed most often. As matter fields do not appear in Eq.(\ref{eq:Omega}) 
for the secondary constraint $\Omega$, there are the same two branches as in the previous subsection
\ber
\Omega_1&=&\omega H_f-\xi H_g =0,\\
\Omega_2&=&\beta_1 \xi^2+2\beta_2 \xi\omega+\beta_3\omega^2=0.
\eer
The second branch  gives nothing new, therefore we consider  the first branch only.
Suppose for simplicity $\kappa=\kappa_g$, then the constraints and dynamical equations are as follows
\ber
\rho_g&=&\frac{2m^2}{\kappa_g}\left(\mu\frac{B_1(r)}{r}-B_0(r)\right),\\
H_g^2&=&\mu\frac{m^2}{3}\frac{B_1(r)}{r},\\
\dot H_g &=&\frac{N}{4}\left[u\kappa_g(\rho_g+p_g)+m^2\left(1-ur\right)D_1(r) \right],\\
\dot H_f &=&-\frac{N}{4}\mu{m^2}(1-ur)\frac{D_1(r)}{r^3},\\
1-ur&=&\frac{(\rho_g+3p_g)+\frac{4m^2}{\kappa_g}\left(\frac{\mu B_1}{r}-B_0 \right)}
{(\rho_g+3p_g)+\frac{2m^2}{\kappa_g}\left(B_0-3D_0 +\frac{3\mu D_1}{r}\right)}.
\eer
Here we cannot freely specify both $\rho_g$, and $H_g$ as the initial data. Only one of these quantities may be fixed, and then we are to solve quartic or cubic equation for $r$. Any variable, besides $N$, can be expressed as a function of $r$, and dynamics of $r$ is given as follows
\be
\dot r=(1-ru)NH_g,
\ee
whereas $N(t)$ is an arbitrary monotonic function.

For an equation of state $p_g=w\rho_g$ we obtain
\be
1-ur=\frac{
(1+w)\left(\frac{\mu B_1}{r}-B_0 \right)
}{
(\frac13+w)\frac{\mu B_1}{r}-(1+w)B_0+\left(r +\frac{\mu }{r}\right)D_1
},
\ee
then the dynamical equations aquire the following form
\ber
\dot r&=&N
\frac{
(1+w)\left(\frac{\mu B_1}{r}-B_0 \right)\sqrt{\mu\frac{m^2}{3}\frac{B_1(r)}{r}}
}
{
(\frac13+w)\frac{\mu B_1}{r}-(1+w)B_0+\left(r +\frac{\mu }{r}\right)D_1},\\
\dot H_g&=&N\frac{m^2}{4}\frac{(1+w)\left[
\left(\frac{2\mu}{3r^2}+3 \right)B_1-\left(\frac{2\mu}{r^2}+3 \right)rD_2
\right]}{
\left(\frac13+w \right)\frac{\mu B_1}{r}-(1+w)B_0+\left(\frac{\mu}{r^2}+1\right)rD_1
}.
\eer


\subsection{One matter minimally interacting with two metrics}
Let us consider the scalar field having a minimal interaction with both $f_{\mu\nu}$ and $g_{\mu\nu}$ metrics~\cite{Akrami:2013ffa,Akrami:2014lja,Yamashita:2014fga,deRham:2014naa}
\be
{\cal L}_\phi=\sqrt{-f}\left(-\frac{1}{2}f^{\mu\nu}\phi_{,\mu}\phi_{,\nu}-{\cal U}(\phi) \right)+
\sqrt{-g}\left(-\frac{1}{2}g^{\mu\nu}\phi_{,\mu}\phi_{,\nu}-{\cal U}(\phi) \right),
\ee
 In the homogeneous cosmology we suppose $\phi=\phi(t)$ and
\be
\mathrm{H}_{matter}={\cal H}_\phi=N\xi^3\rho, \qquad 
\rho=\frac{\pi_\phi^2}{2\xi^6\left(\frac{1}{u}+r^3\right)}+(u+r^3){\cal U}(\phi), 
\ee
Then the Hamiltonian is as follows
\be
\mathrm{H}=\mathrm{H}_{potential}+\mathrm{H}_{kinetic}+\mathrm{H}_{matter}=N{\cal R},
\ee
and the primary constraint-like equations are 
\ber
{\cal R}&=&\xi^3\left[-6\left(\frac{r^3H_f^2}{\kappa_f}+\frac{uH_g^2}{\kappa_g}\right)+\rho(u) +\frac{2m^2}{\kappa}\left(B_1(r)+uB_0(r)\right)\right]=0,\nonumber\\
{\cal S}&\equiv&\frac{\partial{\cal R}}{\partial u}=\xi^3\left[-\frac{6H_g^2}{\kappa_g}+\rho'(u)+\frac{2m^2}{\kappa}B_0(r)\right],
\eer
where 
\be
\rho'(u)=\frac{\partial\rho}{\partial u},
\ee
but both ${\cal R}$ and ${\cal S}$ nonlinearly depend on the auxiliary variable $u$. Then we should solve equation ${\cal S}=0$ for $u$ and substitute the result into ${\cal R}$. Therefore this model contains only one (first class) constraint ${\cal R}=0$ responsible for the time reparametrization invariance, and so it is not free of the Boulware-Deser ghost, as also the model to be considered in Section~\ref{sec:nondRGT}. Here it is possible to specify more initial data than in the previous subsections, for example, $\rho$, $H_g$, and $H_f$.

\subsection{One matter minimally interacting with the effective metric}

The new variant of interaction between the matter and bigravity is by means of the effective metric. When we do not restrict ourselves to mini-superspace 
the effective metric coupling leads to reappearance of the ghost as shown in Refs.~\cite{deRham:2014naa,deRham:2014fha,Noller:2014sta,Soloviev:2014eea,Heisenberg:2014rka}. Nevertheless it is claimed~\cite{deRham:2014naa,deRham:2014fha} that  there is a valid interesting region of applicability for this coupling at scales below the cut-off.
Also, in a recent article~\cite{Hinterbichler:2014} it is stated that the ghost can be excluded in a version of the terad formalism that is non-equivalent to the metric one. 

The first analysis of cosmology in this model was given in article~\cite{Enander:2014xga}.
The effective metric is constructed according to the formula
\be
{\cal G}_{\mu\nu}=\alpha^2g_{\mu\nu}+2\alpha\beta g_{\mu\alpha}\mathsf{X}^\alpha_\nu+\beta^2f_{\mu\nu},\label{eq:eff}
\ee
in tetrad language this corresponds to exploiting the linear combination of the two tetrads:
\be
\alpha E_\mu^A+\beta F_\mu^A,
\ee 
such a possibility for the special case $\alpha=\beta$ has been mentioned first in Ref.~\cite{Krasnov}.
In minisuperspace the components of the effective metric will be denoted as follows
\be
{\cal G}_{00}=-{\cal N}^2,\qquad {\cal G}_{0i}=0,\quad {\cal G}_{ij}=a^2\delta_{ij}\equiv\psi_{ij},
\ee
where
\be
{\cal N}=N(\alpha u+\beta),\qquad a=\alpha \xi+\beta\omega.
\ee
There are the following standard relations:
\be
\sqrt{-{\cal G}}={\cal N}\sqrt{\psi},\qquad \sqrt{\psi}=a^3,
\ee
where
\be
{\cal G}=\det({\cal G}_{\mu\nu}),\qquad \psi=\det(\psi_{ij}).
\ee
The primary constraints are
\ber
{\cal S}&=& -\frac{6{\xi}^3}{\kappa_g}{H_g}^2+\alpha \hat{\cal H}^{(m)}+\frac{2m^2}{\kappa}{\xi}^3B_0(r),\label{eq:S}\\
{\cal R}'&=& -\frac{6\omega^3}{\kappa_f}{H_f}^2+ \beta \hat{\cal H}^{(m)}+\frac{2m^2}{\kappa}\xi^3B_1(r)\label{eq:R},
\eer
where it is supposed that the scalar field canonical variables $(\phi,\pi_\phi)$ depend on time variable only and therefore
\be
\hat{\cal H}^{(m)}=\frac{\pi_\phi^2}{2a^3}+ a^3{\cal U}(\phi).
\ee
We  introduce the following notation for the matter energy density
\be
\rho=\frac{\hat{\cal H}^{(m)}}{a^3}\equiv\frac{\pi_\phi^2}{2a^6}+ {\cal U}(\phi).
\ee
The minisuperspace Hamiltonian is as follows
\be
\mathrm{H}=N({\cal R}'+u{\cal S}),\label{eq:Heff}
\ee
and the Poisson brackets are 
\be
\{\xi,H_g\}=-\frac{\kappa_g}{12\xi^2 },\qquad \{\omega,H_f\}=-\frac{\kappa_f}{12\omega^2 },\qquad \{\phi,\pi_\phi\}=1.\label{eq:PBeff}
\ee
It is suitable to write Eqs.(\ref{eq:S}), (\ref{eq:R}) in a different form
\ber
{\cal S}&=& \xi^3\left[-\frac{6}{\kappa_g}{H_g}^2+\alpha(\alpha +\beta r)^3\rho+\frac{2m^2}{\kappa}B_0(r)\right],\label{eq:SS}\\
{\cal R}'&=&\omega^3\left[ -\frac{6}{\kappa_f}{H_f}^2+ \frac{\beta}{r^3}(\alpha +\beta r)^3\rho +\frac{2m^2}{\kappa}\frac{B_1(r)}{r^3}\right]\label{eq:RR}.
\eer
Demanding the fulfilment of equation ${\cal S}=0$ in the course of evolution we obtain the secondary constraint,
\be
\dot{\cal S}=\{{\cal S},\mathrm{H}\}=N\Omega,\qquad \rightarrow \qquad \Omega=0, 
\ee
where $\Omega$ has the following form 
\be
\Omega=\{{\cal S},{\cal R}'\}=\frac{6m^2}{\kappa}\left(\omega H_f-\xi H_g\right)\left(\beta_1 \xi^2+2\beta_2 \xi\omega+\beta_3\omega^2 -\frac{\kappa}{2m^2}\alpha\beta{a}^2p \right),\label{eq:Omegaeff}
\ee
and we have introduced the notation for pressure

\be
p=
\frac{\pi_\phi^2}{2a^6}-{\cal U}(\phi).
\ee
After calculating 
\be
\{\Omega,{\cal S}\}=\Delta\ne 0,
\ee
we see that $\Omega$ and ${\cal S}$ are 2nd class constraints.
It is possible to introduce the Dirac brackets
\be
\{F,G\}_D=\{F,G\}-\frac{\{F,\Omega\}\{{\cal S},G\}-\{F,{\cal S}\}\{\Omega,G\}}{\Delta},
\ee
and consider  ${\cal R}'$ as a first class contraint. The reduced Hamiltonian and equations of motion will be  as follows
\be
\mathrm{H_{reduced}}=
N{\cal R}', \qquad \dot F=\{F,\mathrm{H_{reduced}}\}_D.
\ee
An equivalent way to derive the Hamiltonian equations is to exploit the Poisson brackets, but after their calculation insert a solution for the variable $u$ determined from the equation
\be
\dot \Omega=\{\Omega,\mathrm{H}\}\equiv N\left(\{\Omega,{\cal R}'\} +u\{\Omega,{\cal S}\}\right)=0,
\ee
i.e.
\be
u=-\frac{1}{\Delta}\{\Omega,{\cal R}'\}.\label{eq:u}
\ee
As $\Omega$ is factorized (\ref{eq:Omegaeff})(compare Eq.(\ref{eq:factor}))
\be
\Omega=\Omega_1\Omega_2,
\ee
there are two different solutions for it:
\be
\Omega_1=\omega H_f-\xi H_g=0, \label{eq:1st2}
\ee
or
\be
\Omega_2=\beta_1\xi^2 +2\beta_2\omega\xi+\beta_3\omega^2-\frac{\kappa}{2m^2}\alpha\beta a^2p=0.\label{eq:2nd2}
\ee
Below we will consider both the  two cases.

With the Hamiltonian (\ref{eq:Heff}) and the Poisson brackets (\ref{eq:PBeff}) we can derive the kinematical Hamiltonian equations, which, of course, are equivalent to the definitions of the Hubble constants (see  (\ref{eq:kinematical}) to compare) and the scalar field momentum
\ber
\dot \xi&=&
\{\xi,H_g\}N\left(\frac{\partial{\cal R}'}{\partial H_g}+u\frac{\partial{\cal S}}{\partial H_g}\right)=Nu\xi H_g,\\
\dot\omega&=&
\{\omega,H_f\}N\left(\frac{\partial{\cal R}'}{\partial H_f}+u\frac{\partial{\cal S}}{\partial H_f}\right)=N\omega H_f,\\
\dot\phi&=&\{\phi,\mathrm{H}\}=N\left(\frac{\partial {\cal R}'}{\partial \pi_\phi}+u\frac{\partial{\cal S}}{\partial\pi_\phi} \right)=N(\alpha u+\beta)\frac{\pi_\phi}{a^3}.
\eer
Combining the kinematical and dynamical Hamiltonian equations for the scalar field (i.e. for the matter)
$$
\dot\phi=N(\alpha u +\beta)\frac{\pi_\phi}{a^3},\qquad \dot\pi_\phi=-N(\alpha u +\beta){a^3}{\cal U}'(\phi)
$$
we obtain the energy conservation law
$$
\frac{\pi_\phi\dot\pi_\phi}{           a^3}+a^3\dot\phi {\cal U}'(\phi)=0,
$$
which can  be written in the standard form
\be
\dot\rho+3\frac{\dot a}{a}(\rho +p)=0.
\ee
Dynamical Hamiltonian equations for the gravitational variables are the following
\ber
\dot H_f&=-&\frac{N\kappa_f}{4r^3}\Biggl[{\beta}\left(\alpha+\beta r\right)^3\left(\rho+p\frac{\beta+\alpha u}{\beta+\frac{\alpha}{r}}\right)+\\
&+&\frac{2m^2}{\kappa}\left(1-ur\right){D_1(r)}\Biggr],\\
\dot H_g&=&-\frac{Nu\kappa_g}{4}\Biggl[\alpha \left(\alpha+\beta r\right)^3\left(\rho+\frac{p}{ur}\frac{\beta+\alpha u}{\beta+\frac{\alpha}{r}}\right)-\\
&-&\frac{2m^2}{\kappa}(1-ur)\frac{D_1(r)}{u}\Biggr].
\eer
(see  (\ref{eq:Hfdot}), (\ref{eq:Hgdot}) for comparison).
Normally the matter equation of state is known also. 

\subsubsection{First branch}
Here we consider the following solution of the constraint $\Omega=0$:
$$
\Omega_1= \omega H_f-\xi H_g=0,\qquad \rightarrow \qquad H_f=r^{-1}H_g.
$$
Then for the Hubble parameters
\be
H_f=\frac{\dot\omega}{N\omega},\qquad H_g=\frac{\dot\xi}{Nu\xi},\qquad H=\frac{\dot a}{N(\alpha u+\beta)a},\label{eq:HHfHg}
\ee
 the following relations are valid
\be
H_g=rH_f, \qquad H_f=H\left(\frac{\alpha}{r}+\beta\right),\qquad H_g=H(\alpha+\beta r).\label{eq:HHHeff}
\ee
Constraints (\ref{eq:SS}), (\ref{eq:RR}) 
are equivalent to equations of the Friedmann form
\ber
H_f^2&=&\frac{\kappa_f}{6r^3}\left(\beta\left(\alpha+\beta r\right)^3\rho+\frac{2m^2}{\kappa}B_1(r)\right),\\
H_g^2&=&\frac{\kappa_g}{6}\left(\alpha(\alpha+\beta r)^3\rho+\frac{2m^2}{\kappa}B_0(r)\right),
\eer 
with a lot of new parameters $\alpha,\beta,\beta_0,\ldots,\beta_4$ and a new variable $r(t)$ involved.
It is suitable to rewrite these equations as follows
\ber
H_f^2&=&\frac{\kappa_f}{6}\rho_f+\frac{\Lambda_f}{3},\\
H_g^2&=&\frac{\kappa_g}{6}\rho_g+\frac{\Lambda_g}{3},
\eer 
where
\ber
\rho_f&=&\beta\left(\frac{\alpha}{r}+\beta
\right)^3\rho,\\
\rho_g&=&\alpha(\alpha+\beta r)^3\rho,\\
\Lambda_f&=&{m^2}\frac{\kappa_f}{\kappa}\frac{B_1(r)}{r^3},\\
\Lambda_g&=&{m^2}\frac{\kappa_g}{\kappa}B_0(r).
\eer 
From Eq.(\ref{eq:u}) we obtain
\be
u=\frac{\frac{B_1}{r}-
3D_2+3\mu^{-1}rD_1+\frac{\kappa}{2m^2}\beta(\alpha+\beta r)^3\left(\rho{r}^{-1}+3p\frac{\beta-\mu^{-1}r\alpha}{\alpha+\beta r}\right)}
{3D_1+\mu^{-1}r(B_0-3D_0)
+\frac{\kappa}{2m^2}\alpha(\alpha+\beta r)^3\left(\rho\mu^{-1}r -3p\frac{\beta-\mu^{-1}r\alpha}{\alpha+\beta r}\right)},
\ee
where
\be
D_i=\beta_i+2\beta_{i+1}r+\beta_{i+2}r^2, \qquad B_i=D_i+rD_{i+1}.
\ee
With account for Eqs.(\ref{eq:HHHeff}) we can reformulate our problem to a couple of Friedmann-like equations for $H$
\ber
H^2&=&\frac{\kappa_g}{6}\alpha(\alpha+\beta r)\rho +\frac{m^2}{3}\frac{\kappa_g}{\kappa}\frac{B_0}{(\alpha+\beta r)^2},\\
H^2&=&\frac{\kappa_f}{6r}\beta\left({\alpha}+\beta{r} \right)\rho+\frac{m^2}{3r}\frac{\kappa_f}{\kappa}\frac{B_1}{(\alpha+\beta r)^2},
\eer
This system can be solved for the matter density   
\be
\rho=\mu^{-1}r\frac{2m^2}{\kappa}
\frac{\frac{\mu B_1(r)}{r}-B_0(r)}
{(\alpha+\beta r)^3\left(\mu^{-1}r{\alpha}-{\beta}\right)}.\label{eq:rhoeff}
\ee
Vice versa it is possible to exclude $\rho$, and obtain
\be
H^2=\frac{m^2}{3}\frac{
{\alpha}B_1(r)-{\beta}B_0(r)
}{
(\alpha+\beta r)^2\left( \mu^{-1}r{\alpha}-{\beta}
\right)
}\frac{\kappa_g}{\kappa},\label{eq:H2r}
\ee
or we can express the cosmological term as a function of $r$:
\be
\Lambda=m^2\frac{\kappa_g}{\kappa}\frac{B_0}{(\alpha+\beta r)^2}.\label{eq:Lambdaeff}
\ee

When the unobservable variable $r$ is given, we can calculte $\rho$, $H$, and $\Lambda$ from Eqs.(\ref{eq:rhoeff})--(\ref{eq:Lambdaeff}). Vice versa, if we take an initial value for $\rho$ or $H$, after solving a quartic or cubic equation we can find a few solutions for $r$. The dynamical equation for $r$
\be
\dot r=\dot a (1-ur)\frac{{\alpha}+\beta{r}}{\beta +\alpha{u}},
\ee
is sufficient to predict its evolution, and therefore the evolution (forward or backward in time) of all observables.
For the important factor $1-ur$ we get the following expression
\be
1-ur=\frac{
\alpha (\alpha+\beta r)^3(\rho+3p)\left(1-\frac{\beta\mu}{\alpha r}\right)+\frac{4m^2}{\kappa}\left(\frac{\mu B_1}{r}-B_0 \right)
}{
\alpha(\alpha+\beta r)^3\left(\rho-3p\frac{\beta-\mu^{-1}r\alpha}{\alpha+\beta r} \right)+\frac{2m^2}{\kappa}\left( B_0-3D_0+\frac{3\mu D_1}{r}\right)
}.\label{eq:1ureff}
\ee
Eq.(\ref{eq:1ureff}) may be simplified if the matter has the equation of state $p=w \rho$:
\be
1-ur=\frac{
3(1+w)\left(\frac{\mu B_1}{r}-B_0 \right)
}{
B_0-3D_0+\frac{3\mu D_1}{r}+\left(\frac{\mu B_1}{r}-B_0 \right)\left(\frac{1}{1-\frac{\mu\beta}{\alpha r}}+\frac{3w}{1+\frac{\beta}{\alpha} r}\right)
}.
\ee
$r(t)$ may be non-monotonic and may change its behavior at those points where $r=1/u$. 

\subsubsection{Second branch}

The second branch $\Omega_2=0$ gives the following equation
$$
\beta_1 \xi^2+2\beta_2 \xi\omega+\beta_3\omega^2 -\frac{\kappa}{2m^2}\alpha\beta {a}^2p =0,
$$
or, equivalently,
$$
p=\frac{2m^2}{\kappa}\frac{D_1(r)}{\alpha\beta(\alpha+\beta r)^2}.
$$
This equation gives a possibility to determine the pressure of matter $p$, and also the density $\rho$, through the matter equation of state, as a function of variable $r$. Then from constraints ${\cal S}=0$, ${\cal R}'=0$ we can find $H_g$ and $H_f$. 
 Evolution of these variables are provided by the Hamiltonian equations
\ber
\dot H_g&=&-\frac{Nu\kappa_g}{4}\alpha(\alpha+\beta r)^3(\rho+p),\label{eq:dotHg}\\
\dot H_f&=&-\frac{N\kappa_f}{4r^3}\beta(\alpha+\beta r)^3(\rho+p).\label{eq:dotHf}
\eer 
The combination of Eqs.(\ref{eq:dotHg}), (\ref{eq:dotHf}) gives the following relation
\be
\frac{\dot H_g}{\dot H_f}=\mu^{-1}\frac{\alpha}{\beta}ur^3.
\ee

\section{Non-dRGT bigravity}\label{sec:nondRGT}
Let the potential in bigravity
will be of non-dRGT type, for example, as taken from the Relativistic Theory of Gravitation (RTG)~\cite{Logunov}:
\be
U=\frac{1}{2}\left[\sqrt{-g}\left(\frac{1}{2}g^{\mu\nu}f_{\mu\nu}-1\right)-\sqrt{-f} \right].
\ee 
Then in $3+1$ minisuperspace notations 
\be 
\sqrt{-g}=Nu\xi^3,\qquad \sqrt{-f}=N\omega^3,\qquad g^{\mu\nu}f_{\mu\nu}=\frac{1}{u^2}+3r^2,
\ee
\be
U=\frac{1}{2} N\left(\frac{\xi^3}{2u}+\frac{3}{2}u\xi\omega^2-u\xi^3-\omega^3\right)\equiv N\tilde U,
\ee
and we obtain
\ber
V&=&\frac{\partial\tilde U}{\partial u}=\frac{1}{2} \left(-\frac{\xi^3}{2u^2}-\xi^3+\frac{3}{2}\xi\omega^2\right),\\
W&=&\tilde U-u\frac{\partial\tilde U}{\partial u}=\frac{1}{2}\left(\frac{\xi^3}{u}-\omega^3\right).
\eer
Therefore constraint-like equations are the following
\ber
{\cal R}'&=&{\cal H}+\frac{2m^2}{\kappa}W=\omega^3\left[-\frac{6H_f^2}{\kappa_f}+\rho_f+\frac{m^2}{\kappa}\left(\frac{1}{ur^3}-1 \right)\right]=0,\label{eq:biRTG1}\\
{\cal S}&=&\bar{\cal H}+\frac{2m^2}{\kappa}V=\xi^3\left[-\frac{6H_g^2}{\kappa_g}+\rho_g+\frac{m^2}{\kappa}\left(-\frac{1}{2u^2}-1+\frac{3}{2}r^2 \right)\right]=0,\label{eq:biRTG2}
\eer
but in fact both Eqs.(\ref{eq:biRTG1}), (\ref{eq:biRTG2}) contain auxiliary variable $u$. Then we should solve one of these equations for $u$ and substitute the solution into another equation. So, there is only one constraint in fact. It is a first class constraint arising as a result of the time reparametrisation invariance having place in the bi-RTG. For this reason the variable $N$ stays arbitrary.  One can take, for example, $\rho_g$, $\rho_f$, $H_g$, and $H_f$ as the initial data. We see that this theory has one more gravitational degree of freedom than the bigravity with the dRGT potential. As a result, it is not suitable for the decription of the observable cosmology, and so never has been discussed in the literature. 

The dynamical Hamiltonian equations here are as follows
\ber
\dot {H_f}&=&-\frac{N\kappa_f}{4}\left[\rho_f+p_f+\frac{m^2}{\kappa}\frac{1-(ur)^2}{ur^3}  \right],\\
\dot {H_g}&=&-\frac{Nu\kappa_g}{4}\left[\rho_g+p_g-\frac{m^2}{\kappa}\frac{1-(ur)^2}{u^2}  \right].
\eer

\subsection{Non-dRGT massive gravity}
When we take one metric, for example, $f_{\mu\nu}$, as  a background Minkowskian one, 
we put $\omega\equiv 1$, $H_f=0$, $\rho_f=0$, and the solution of Eq.(\ref{eq:biRTG1}) is
\be
u=\frac{1}{r^3}\equiv\frac{\xi^3}{\omega^3}=\xi^3.
\ee
After substitution of this into ${\cal S}$ we obtain the Friedmann equation in the RTG:
\be
H_g^2=\frac{\kappa_g\rho_g}{6}-\frac{m^2}{12\xi^6}\frac{\kappa_g}{\kappa}\left( 1+2\xi^6-3\xi^4\right).\label{eq:RTG}
\ee
If we take $\kappa=\kappa_g=16\pi G$, then it is possible to rewrite Eq.(\ref{eq:RTG}) as follows
\be
H_g^2=
\frac{8\pi G}{3}\rho_g+\frac{\Lambda}{3},\label{eq:RTG2}
\ee
where
\be
\Lambda(\xi)=-\frac{m^2}{2}\left(
 \frac{1}{2\xi^6}+1-\frac{3}{2\xi^2}
\right).\label{eq:RTG3}
\ee
We can take  $\rho_g$ and $H_g$ as the initial data. Then $\Lambda(\xi)$ is determined from Eq.(\ref{eq:RTG2}), and $\xi$ could be found as a solution of the bi-cubic equation (\ref{eq:RTG3}).
In contrast to the massive gravity with dRGT potential this theory admits the homogeneous and isotropic dynamical Friedmann cosmology. It is due to the fact that the gravitational field has one additional degree of freedom here. But this degree of freedom is unfortunately a ghost. 
The effective cosmological constant is negative, and therefore any expansion should in future be transformed into contraction. The fate of the Universe is cyclic in this model. Of course, a  quintessence matter field may provide an accelerated expansion for a finite interval of time.

\section{Conclusion}
The 
cosmology of bigravity 
is an open problem that is under active study~\cite{Comelli:2015pua,Gumrukcuoglu:2015,25March2015,Sakakihara2015}. 
This article is written in order to demonstrate the beauty and power of the Hamiltonian formalism 
in this field of research. 
We would like to mention that it is a direct road to develop the quantum cosmology in bigravity. 

The author is grateful to V. A. Petrov, Yu. F. Pirogov, and Yu. M. Zinoviev
for the interest to this work.

\section{Appendix}
The straightforward calculation of traces for matrix $\mathsf{X}$ starting from Eq.(\ref{eq:ABCD}) gives the following results
\begin{eqnarray}
\mathrm{Tr}\mathsf{X}&=&-A+D,\\
\mathrm{Tr}\mathsf{X}^2&=&A^2-2(BC)+\mathrm{Tr}D^2,\\
\mathrm{Tr}\mathsf{X}^3&=&-A^3+3A(BC)-(BDC)-\mathrm{Tr}D^3.
\end{eqnarray}
Then it is easy to estimate the function $\tilde U$ and its derivatives
\begin{eqnarray}
V&=&\beta_0e+\beta_1e\left(x+\frac{y}{\varepsilon+1}\right)+\beta_2e\left[x^2-\mathrm{Tr}x^2+2\frac{xy-(px^2p)}{\varepsilon +1} \right]+\nonumber \\
&+&\beta_3e\left[ x^3-3x\mathrm{Tr}x^2+2\mathrm{Tr}x^3+\frac{6(px^3p)-6x(px^2p)-3y\mathrm{Tr}x^2}{\varepsilon+1}\right]\nonumber\\
V_i&=&-\beta_1e(fp)^i+2\beta_2e\left[(fx^{\mathrm{T}}p)^i-(fp)^ix \right]
+ \nonumber\\
&+&3\beta_3e\left[2x(fx^{\mathrm{T}}p)^i+\mathrm{Tr}x^2(fp)^i-2(f(x^2)^{\mathrm{T}}p)^i \right]=- {\bf f}_{ia}C_{ab}p_b,\nonumber\\
W&=&\beta_4f+\beta_1e\varepsilon+2\beta_2e\varepsilon\left( x-\frac{y}{\varepsilon(\varepsilon+1)} \right)+\nonumber\\
&+&\beta_3e\left[ x^3-3x\mathrm{Tr}x^2+2\mathrm{Tr}x^3+\frac{6(px^3p)-6x(px^2p)-3z\mathrm{Tr}x^2}{\varepsilon+1}\right],\nonumber
\end{eqnarray}
where
\be
C_{ab}=e\left[\delta_{ab}(\beta_1+2\beta_2x-3\beta_3{\mathrm{Tr}x^2})-2x_{ba}(\beta_2+3\beta_3x)+6\beta_3x_{bc}x_{ca} \right].
\ee
The following notations were used above
\begin{eqnarray}
x_{ab}&=&{\bf f}_{ia}{\bf e}^i_b,\\
x&=&x_{aa},\\
\mathrm{Tr}x^2&=&x_{ab}x_{ba},\\
\mathrm{Tr}x^3&=&x_{ab}x_{bc}x_{ca},\\
y_{ab}&=&p_ap_c x_{cb},\\
y&\equiv&(pxp)=p_ax_{ab}p_b=y_{aa},\\
(px^2p)&=&p_ax_{ab}x_{bc}p_c,\\
(px^3p)&=&p_ax_{ab}x_{bc}x_{cd}p_d,\\
(ufp)&=&u^if_{ia}p^a,\\
(ufx^{\mathrm{T}}p)&=&u^i f_{ia}x_{ba}p_b,\\
(uf({x^2})^{\mathrm{T}}p)&=&u^i f_{ia}x_{ba}x_{cb}p_c.
\end{eqnarray}


\begin{thebibliography}{*}
\bibitem{Boulware:1973my}
D.~Boulware and S.~Deser, {\it {Can gravitation have a finite range?}},  {\sl
  Phys.Rev.} {\bf D6} (1972) 3368--3382,  [\href{http://dx.doi.org/10.1103/physrevd.6.3368}{{\sf
  doi:10.1103/physrevd.6.3368}}].  

\bibitem{deRham:2010ik}
C.~de~Rham and G.~Gabadadze, {\it {Generalization of the Fierz-Pauli action}},
  {\sl Phys.Rev.} {\bf D82} (2010) 044020,
  [\href{http://arxiv.org/abs/1007.0443}{{\sf arXiv:1007.0443}}].

\bibitem{deRham:2010kj}
C.~de~Rham, G.~Gabadadze, and A.~J. Tolley, {\it {Resummation of massive
  gravity}},  {\sl Phys.Rev.Lett.} {\bf 106} (2011) 231101,
  [\href{http://arxiv.org/abs/1011.1232}{{\sf arXiv:1011.1232}}],  [\href{http://dx.doi.org/10.1103/physrevlett.106.231101}{{\sf
  doi:10.1103/physrevlett.106.231101}}].  

\bibitem{HR_bi}
	 	S.~F. Hassan, R.~A. Rosen,
	 	{\it Bimetric Gravity from Ghost-free Massive Gravity},
{\sl JHEP}
{\bf 02} 126 (2012),	 	 	
[\href{http://arxiv.org/abs/1109.3515}{{\sf arXiv:1109.3515}}], [\href{http://dx.doi.org/10.1007/jhep02(2012)126 }{{\sf
  doi:10.1007/jhep02(2012)126}}]. 



\bibitem{HR2}
S.~F. Hassan, R.~A. Rosen, 
	{\it Confirmation of the Secondary Constraint and Absence of Ghost in Massive   Gravity and Bimetric Gravity}, 
{\sl JHEP} {\bf 04}  123 (2012), [\href{http://arxiv.org/abs/1111.2070}{{\sf arXiv:1111.2070}}], [\href{http://dx.doi.org/10.1007/jhep04(2012)123 }{{\sf
  doi:10.1007/jhep04(2012)123}}]. 



\bibitem{Comelli:2015pua}
D.~Comelli, M.~Crisostomi, K.~Koyama, L.~Pilo, and G.~Tasinato, {\it {Cosmology
  of bigravity with doubly coupled matter}}, {\sl JCAP} {\bf 04} 026 (2015),
  [\href{http://arxiv.org/abs/1501.00864}{{\sf arXiv:1501.00864}}],  	
[\href{http://dx.doi.org/10.1088/1475-7516/2015/04/026
}{{\sf  doi:10.1088/1475-7516/2015/04/026}}]
 	
\bibitem{Gumrukcuoglu:2015}
A. Emir Gumrukcuoglu, Lavinia Heisenberg, Shinji Mukohyama, Norihiro Tanahashi,
{\it Cosmology in bimetric theory with an effective composite coupling to   matter},
[\href{http://arxiv.org/abs/1501.02790}{{\sf arXiv:1501.02790}}].

\bibitem{25March2015}
Yashar Akrami, S. F. Hassan, Frank K{\"o}nnig, Angnis Schmidt-May, Adam R. Solomon,
{\it Bimetric gravity is cosmologically viable},
 [\href{http://arxiv.org/abs/1503.07521}{{\sf arXiv:1503.07521}}].

\bibitem{SolTch}
V. O. Soloviev, M. V. Tchichikina, {\it Bigravity in Kucha\u{r}'s Hamiltonian formalism. The general case},  {\it Theor. Math. Phys.} {\bf 176} 1163-1175 (2013),
[\href{http://arxiv.org/abs/1211.6530}{{\sf arXiv:1211.6530}}], [\href{http://dx.doi.org/10.1007\%2Fs11232-013-0097-y}{{\sf doi:10.1007/TMPh176(2013)1163}}].

\bibitem{SolTch2}
V. O. Soloviev, M. V. Tchichikina, {\it Bigravity in Kucha\u{r}'s Hamiltonian formalism: The special case}, 
{\sl Phys. Rev.} {\bf D 88} 084026 (2013), 
[\href{http://arxiv.org/abs/1302.5096}{{\sf arXiv:1302.5096}}], [\href{http://dx.doi.org/10.1103/PhysRevD.88.084026}{{\sf
  doi:10.1103/PhysRevD.88.084026}}].

\bibitem{Comelli}
D. Comelli, F. Nesti, L. Pilo, Weak massive gravity, {\sl Phys. Rev.} {\bf D 87} 124021 (2013),    [\href{http://dx.doi.org/10.1103/PhysRevD.87.124021}{{\sf
  doi:10.1103/PhysRevD.87.124021}}], \href{http://arxiv.org/abs/1302.4447}{{\sf arXiv:1302.4447}};
 Massive gravity: a General Analysis, {\it JHEP} {\bf 2013} 2013:161, 
 [\href{http://dx.doi.org/10.1007/JHEP07(2013)161}{{\sf
  doi:10.1007/JHEP07(2013)161}}],
\href{http://arxiv.org/abs/1305.0236}{{\sf arXiv:1305.0236}}.

\bibitem{Sol}
V. O. Soloviev, 
{\it Bigravity in Hamiltonian formalism: The tetrad approach}, 
{\sl Theor. Math. Phys.} 
{\bf 182} 
294-307 (2015)  
[\href{http://dx.doi.org/10.1007\%2Fs11232-015-0263-5}{{\sf
  doi:10.1007/TMPh182(2015)294}}].

\bibitem{Soloviev:2014eea}
V.~O. Soloviev, {\it {Bigravity in tetrad Hamiltonian formalism and matter  couplings}},  
\href{http://arxiv.org/abs/1410.0048}{{\sf arXiv:1410.0048}}.

\bibitem{Kluson_tetrad}
J. Kluson,
{\it Hamiltonian Formalism of Bimetric Gravity In Vierbein Formulation}, Eur. Phys. J. C, vol. 74, no. 8, 2985,
\href{http://arxiv.org/abs/1307.1974}{{\sf arXiv:1307.1974}},   [\href{http://dx.doi.org/10.1007/s10714-013-1639-1}{{\sf
  doi:10.1140/epjc/s10052-014-2985-1}}]. 

\bibitem{Golovnev}
Alexey Golovnev,
{\it On the Hamiltonian analysis of non-linear massive gravity}, {\sl Phys. Lett.} {\bf  B 707} 404-408  (2012),  
[\href{http://arxiv.org/abs/1112.2134}{{\sf arXiv:1112.2134}}],  [\href{http://dx.doi.org/10.1016/j.physletb.2011.12.064}{{\sf
  doi:10.1016/j.physletb.2011.12.064}}]. 
    
\bibitem{Deser}
S. Deser, M. Sandora, A. Waldron, G. Zahariade,
{\it    Covariant constraints for generic massive gravity and analysis of its characteristics},
{\sl Phys. Rev.} {\bf D 90} 104043 (2014),
 \href{http://arxiv.org/abs/1408.0561}{{\sf arXiv:1408.0561}}, 
[\href{http://dx.doi.org/10.1103/PhysRevD.90.104043}{{\sf
  doi:10.1103/PhysRevD.90.104043}}].
  


\bibitem{Ryan}
M. Ryan, {\it Hamiltonian Cosmology}, Springer-Verlag, Berlin-Heidelberg-New York, (1972).
Lecture Notes in Physics, Vol. 13, Eds. J. Ehlers, K. Hepp, H.A. Weidenmueller.





\bibitem{Kuch1}
K. Kucha\u{r}, 
{\it Geometry of hyperspace. I}, 
{\sl J. Math. Phys.}, 1976, vol. 17, no. 5, pp. 777--791,
 [\href{http://dx.doi.org/10.1063/1.522976}{{\sf
  doi:10.1063/1.522976}}].


\bibitem{Kuch2}
K. Kucha\u{r}, 
{\it Kinematics of tensor fields in hyperspace. II}, 
{\sl J. Math. Phys.}, 1976, vol. 17, no. 5,
pp. 792--800, 
 [\href{http://dx.doi.org/10.1063/1.522977}{{\sf
  doi:10.1063/1.522977}}].


\bibitem{Kuch3}
K. Kucha\u{r}, 
{\it Dynamics of tensor fields in hyperspace. III}, 
{\sl J. Math. Phys.}, 1976, vol. 17, no. 5,
pp. 801--820,
 [\href{http://dx.doi.org/10.1063/1.522978}{{\sf
  doi:10.1103/1.522978}}].


\bibitem{Kuch4}
K. Kucha\u{r}, 
{\it Geometrodynamics with tensor sources. IV}, 
J. Math. Phys., 1977, vol. 18, no. 8,
pp. 1589--1597, 
 [\href{http://dx.doi.org/10.1063/1.523467}{{\sf
  doi:10.1103/1.523467}}].





\bibitem{Dirac}
P. A. M. Dirac,  {\it Lectures on Quantum Mechanics}, Yeshiva University, New York, (1964). 

\bibitem{Leznov}
D. Fairlie, A. Leznov,
{\it General solutions of the Monge-Amp\`ere equation in $n$-dimensional space},
{\sl J. Geom. Phys.} {\bf 16} 385-390 (1995);   \href{http://arxiv.org/abs/hep-th/9403134}{{\sf arXiv:hep-th/9403134}}.

\bibitem{HamTetrads}
S. Deser and C. J. Isham. {\sl Phys. Rev.} {\bf D 14} 2505-2510 (1976); J.E. Nelson and C. Teitelboim. {\sl Annals of Physics} {\bf 116} 86-104 (1978); M. Henneaux. {\sl Gen. Rel. Grav.} {\bf 9} 1031-1045 (1978).

\bibitem{Krasnov}
S.~Alexandrov, K.~Krasnov, and S.~Speziale,
{\it Chiral description of ghost-free massive gravity}, JHEP 1306 (2013) 068,
\href{http://arxiv.org/abs/1212.3614}{{\sf arXiv:1212.3614}}, [\href{http://dx.doi.org/10.1007/JHEP06(2013)068}{{\sf
  doi:10.1007/JHEP06(2013)068}}]. 



\bibitem{Alexandrov}
S. Alexandrov,
Canonical structure of Tetrad Bimetric Gravity, Gen.Rel.Grav. 46 (2014), 
\href{http://arxiv.org/abs/1308.6586}{{\sf arXiv:1308.6586}}, [\href{http://dx.doi.org/10.1007/s10714-013-1639-1}{{\sf
  doi:1639.10.1007/s10714-013-1639-1}}]. 


\bibitem{Sakakihara2015}
Yuki Sakakihara, Jiro Soda,
{\it Primordial Gravitational Waves in Bimetric Gravity},
[\href{http://arxiv.org/abs/1504.04969}{{\sf arXiv:1504.04969}}].

\bibitem{Volkov:2011an}
M.~S. Volkov, {\it {Cosmological solutions with massive gravitons in the
  bigravity theory}},  {\sl JHEP} {\bf 1201} (2012) 035,
  [\href{http://arxiv.org/abs/1110.6153}{{\sf arXiv:1110.6153}}],
  [\href{http://dx.doi.org/10.1007/JHEP01(2012)035}{{\sf
  doi:10.1007/JHEP01(2012)035}}].

\bibitem{vonStrauss:2011mq}
M.~von Strauss, A.~Schmidt-May, J.~Enander, E.~Mortsell, and S.~Hassan, {\it
  {Cosmological Solutions in Bimetric Gravity and their Observational Tests}},
  {\sl JCAP} {\bf 1203} (2012) 042, [\href{http://arxiv.org/abs/1111.1655}{{\sf
  arXiv:1111.1655}}],
  [\href{http://dx.doi.org/10.1088/1475-7516/2012/03/042}{{\sf
  doi:10.1088/1475-7516/2012/03/042}}].

\bibitem{Comelli:2011zm}
D.~Comelli, M.~Crisostomi, F.~Nesti, and L.~Pilo, {\it {FRW Cosmology in Ghost
  Free Massive Gravity}},  {\sl JHEP} {\bf 1203} (2012) 067,
  [\href{http://arxiv.org/abs/1111.1983}{{\sf arXiv:1111.1983}}],
  [\href{http://dx.doi.org/10.1007/JHEP06(2012)020,
  10.1007/JHEP03(2012)067}{{\sf doi:10.1007/JHEP06(2012)020,
  10.1007/JHEP03(2012)067}}].

\bibitem{Akrami:2012vf}
Y.~Akrami, T.~S. Koivisto, and M.~Sandstad, {\it {Accelerated expansion from
  ghost-free bigravity: a statistical analysis with improved generality}},
  {\sl JHEP} {\bf 1303} (2013) 099, [\href{http://arxiv.org/abs/1209.0457}{{\sf
  arXiv:1209.0457}}], [\href{http://dx.doi.org/10.1007/JHEP03(2013)099}{{\sf
  doi:10.1007/JHEP03(2013)099}}].



\bibitem{PhysRevD.84.124046}
G.~D'Amico, C.~de~Rham, S.~Dubovsky, G.~Gabadadze, D.~Pirtskhalava, and A.~J.
  Tolley, {\it Massive cosmologies},  {\sl Phys. Rev. D} {\bf 84} (Dec, 2011)
  124046, [\href{http://dx.doi.org/10.1103/PhysRevD.84.124046}{{\sf
  doi:10.1103/PhysRevD.84.124046}}].

\bibitem{Gumrukcuoglu:2011ew}
A.~E. Gumrukcuoglu, C.~Lin, and S.~Mukohyama, {\it {Open FRW universes and
  self-acceleration from nonlinear massive gravity}},  {\sl JCAP} {\bf 1111}
  (2011) 030, [\href{http://arxiv.org/abs/1109.3845}{{\sf arXiv:1109.3845}}].

\bibitem{Akrami:2013ffa}
Y.~Akrami, T.~S. Koivisto, D.~F. Mota, and M.~Sandstad, {\it {Bimetric gravity
  doubly coupled to matter: theory and cosmological implications}},  {\sl JCAP}
  {\bf 1310} (2013) 046, [\href{http://arxiv.org/abs/1306.0004}{{\sf
  arXiv:1306.0004}}],
  [\href{http://dx.doi.org/10.1088/1475-7516/2013/10/046}{{\sf
  doi:10.1088/1475-7516/2013/10/046}}].



\bibitem{Akrami:2014lja}
Y.~Akrami, T.~S. Koivisto, and A.~R. Solomon, {\it {The nature of spacetime in
  bigravity: two metrics or none?}},
  \href{http://arxiv.org/abs/1404.0006}{{\sf arXiv:1404.0006}}.



\bibitem{Yamashita:2014fga}
Y.~Yamashita, A.~De~Felice, and T.~Tanaka, {\it {Appearance of Boulware-Deser
  ghost in bigravity with doubly coupled matter}},
  \href{http://arxiv.org/abs/1408.0487}{{\sf arXiv:1408.0487}}.

\bibitem{deRham:2014naa}
C.~de~Rham, L.~Heisenberg, and R.~H. Ribeiro, {\it {On couplings to matter in
  massive (bi-)gravity}},  \href{http://arxiv.org/abs/1408.1678}{{\sf
  arXiv:1408.1678}}.

\bibitem{deRham:2014fha}
C.~de~Rham, L.~Heisenberg, and R.~H. Ribeiro, {\it {Ghosts and Matter Couplings
  in Massive (bi-and multi-)Gravity}},  {\sl Phys.Rev.} {\bf D90} (2014)
  124042, [\href{http://arxiv.org/abs/1409.3834}{{\sf arXiv:1409.3834}}],
  [\href{http://dx.doi.org/10.1103/PhysRevD.90.124042}{{\sf
  doi:10.1103/PhysRevD.90.124042}}].

\bibitem{Noller:2014sta}
J.~Noller and S.~Melville, {\it {The coupling to matter in Massive, Bi- and
  Multi-Gravity}},  \href{http://arxiv.org/abs/1408.5131}{{\sf
  arXiv:1408.5131}}.



\bibitem{Heisenberg:2014rka}
L.~Heisenberg, {\it {Quantum corrections in massive bigravity and new effective
  composite metrics}},  \href{http://arxiv.org/abs/1410.4239}{{\sf
 arXiv:1410.4239}}.

\bibitem{Hinterbichler:2014}
K. Hinterbichler, and R. A. Rosen,
{\it A Note on Ghost-Free Matter Couplings in Massive Gravity and Multi-Gravity},
\href{http://arxiv.org/abs/1503.06796/}{{\sf arXiv:1503.06796}}.

\bibitem{Enander:2014xga}
J.~Enander, A.~R. Solomon, Y.~Akrami, and E.~Mortsell, {\it {Cosmic expansion
  histories in massive bigravity with symmetric matter coupling}},
  \href{http://arxiv.org/abs/1409.2860}{{\sf arXiv:1409.2860}}.

\bibitem{Logunov}
S. S. Gershtein, A. A. Logunov, and M. A. Mestvirishvili,
{\it Upper limit on the graviton mass}, \href{http://arxiv.org/abs/hep-th/9711147}{{\sf arXiv:hep-th/9711147}}.


\bibitem{DM}
Luc Blanchet, Lavinia Heisenberg,
Dark Matter via Massive (bi-) Gravity, \href{http://arxiv.org/abs/1504.00870}{{\sf arXiv:1504.00870}}.


\end{thebibliography}
\end{document}